\documentclass[twocolumn,prx,aps,superscriptaddress]{revtex4-2}
\usepackage{amsmath}
\usepackage{amssymb}
\usepackage{amsfonts}
\usepackage[dvips]{graphicx}
\usepackage{subfigure}
\usepackage{dcolumn}
\usepackage{txfonts}
\usepackage{bm}
\usepackage{makeidx}
\usepackage{color}
\usepackage{mathtools}
\usepackage{threeparttable}
\usepackage[colorlinks,linkcolor=blue,anchorcolor=blue,citecolor=blue,urlcolor=blue]{hyperref}

\begin{document}

\preprint{APS/123-QED}

\title{Quantum structured light: Non-classical spin texture of twisted single-photon pulses}
\author{Li-Ping Yang}

\affiliation{Center for Quantum Sciences and School of Physics, Northeast Normal University, Changchun 130024, China}

\affiliation{Birck Nanotechnology Center, School of Electrical and Computer Engineering, Purdue University, West Lafayette, IN 47906, U.S.A.}

\author{Zubin Jacob}
\email{zjacob@purdue.edu}
\affiliation{Birck Nanotechnology Center, School of Electrical and Computer Engineering, Purdue University, West Lafayette, IN 47906, U.S.A.}

\maketitle
\textbf{Abstract}

Classical structured light with controlled polarization and orbital angular momentum (OAM) of electromagnetic waves has varied applications in optical trapping, bio-sensing, optical communications and quantum simulations. However, quantum noise and photon statistics of three-dimensional photonic angular momentum are relatively less explored. Here, we develop a quantum framework and put forth the concept of quantum structured light for space-time wavepackets at the single-photon level. Our work deals with three-dimensional angular momentum observables for twisted quantum pulses beyond scalar-field theory as well as the paraxial approximation. We show that the spin density generates modulated helical texture and exhibits distinct photon statistics for Fock-state vs. coherent-state twisted pulses. We introduce the quantum correlator of photon spin density to characterize nonlocal spin noise providing a rigorous parallel with electronic spin noise. Our work can lead to quantum spin-OAM physics in twisted single-photon pulses and opens explorations for phases of light with long-range spin order.

\textbf{Introduction}
Structured single-photon pulses are an important frontier for spin and orbital angular momentum (OAM)~\cite{Forbes2021Structured,Forbes2019quantum,Erhard2018twisted}. As a quantum information carrier, single-photon pulses with OAM have been achieved in the solid-state system with quantum dots recently~\cite{Chen2021bright} and have been exploited to construct a quantum network with higher channel capacity~\cite{krenn2015twisted,zhang2020tunable,Sroor2020high,Ding2015storage,Zhou2015Storage,Malik2016multiphoton}. The spin and OAM of light have also attracted increasing attention in an emerging research field---spin-orbit photonics~\cite{cardano2015spin}, which studies photon spin-OAM transfer~\cite{Fang2021photo,devlin2017arbitrary,stav2018quantum,aiello2015transverse} and light-matter angular momentum exchange in the near-field region~\cite{rodriguez2013near,petersen2014chiral,gong2018nanoscale} or transfer of optical OAM to bounded electrons~\cite{schmiegelow2016transfer} or photoelectrons~\cite{Fang2021photo,Matula2013Atomic}. Spin-1 quantization is also the hallmark of photonic skyrmions and topological photonic phases of matter ~\cite{barik2018topological,mechelen2020viscous}. A  quantum field theory framework is needed to study the non-classical properties such as 3D noise of the angular momentum of light.  

\begin{figure*}
\centering
\includegraphics[width=16cm]{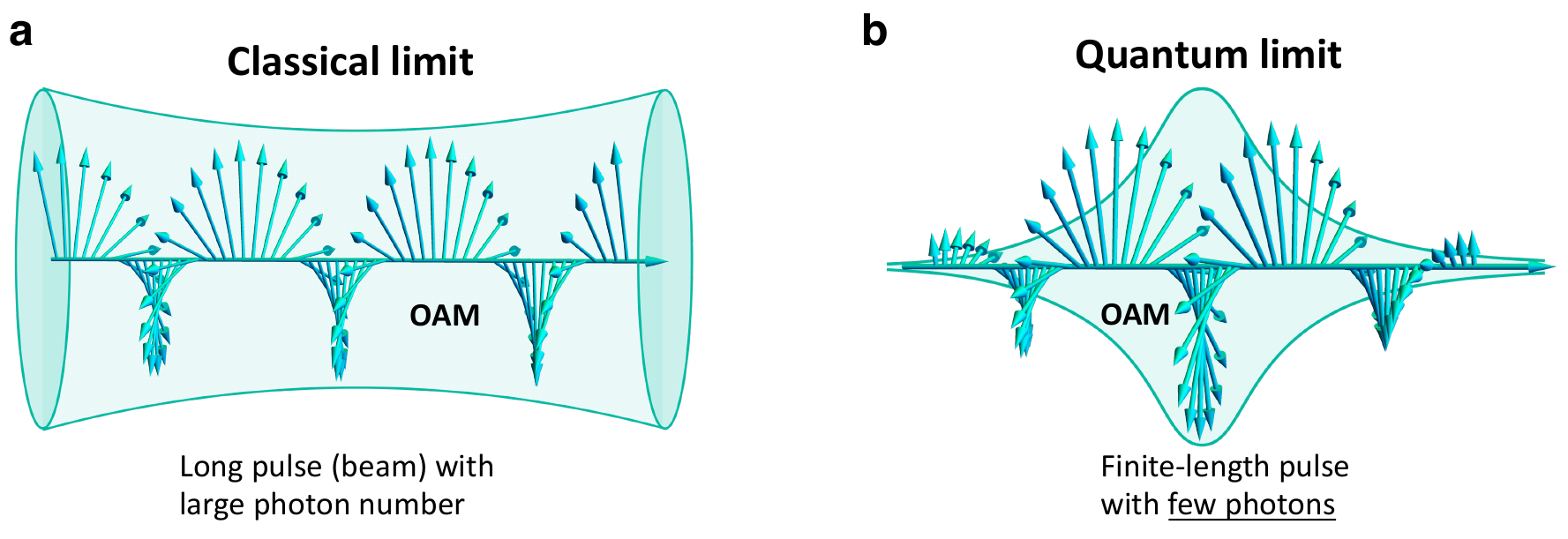}\caption{\label{fig:schematic} \textbf{Contrast between traditional twisted beam and quantum pulse.} Schematic of a traditional twisted beam \textbf{a} compared to the quantum twisted pulse \textbf{b} put forth in this paper. The semi-classical theory only captures the mean orbital angular momentum (OAM) of a laser beam with large photon number. However, the quantum effects of photon statistics, vectorial uncertainty relations and non-local spin noise requires a quantum theoretical framework put forth in this paper.}
\end{figure*}

Existing theories of quantum light-matter interaction have advanced over the last two decades to capture a plethora of phenomena related to SAM and OAM of light ~\cite{barnett1994orbital,berry1998paraxial,Monteiro2009angular,Li2009spin,cerjan2011orbital,Holleczek2011classical,bliokh2015transverse,Arnaut2000orbital,Jauregui2005quantum,Calvo2006quantum,Milione2011higher,li2020spin}. Important outstanding questions remain even within this large body of work which is the focus of this manuscript, namely - photon statistics, 3D quantum spin and OAM vector density, 3D quantum noise in SAM/OAM and single photon quantum states. Fig.~\ref{fig:schematic}\textbf{a} shows the well-known regime of twisted laser beams which contain an enormous number of photons.  At the single photon level, both existing semi-classical~\cite{barnett1994orbital,berry1998paraxial} and approximate quantum theories break down~\cite{Calvo2006quantum,Arnaut2000orbital}. In the widely adopted  state-space description of single photons or entangled photons $\{|l,s\rangle\}$ with Stokes parameters and the Poincar{\' e} sphere~\cite{Milione2011higher,Holleczek2011classical},  the rich spatial texture of spin and OAM vectors are ignored completely. Specifically, important open questions remain on the full 3D projection of photon spin and OAM at the quantum level beyond the scalar-field theory and paraxial approximation. Heisenberg uncertainty relations for photon angular momenta can affect quantum metrology experiments which require a quantum theoretic framework. \textcolor{red}{For the observable photon OAM, these uncertainty relations must account for its noncanonical spin--orbital commutation algebra.}   Similarly, for applications such as secure quantum communication,  twisted single-photon pulses in the quantum limit with few photons (see Fig.~\ref{fig:schematic}\textbf{b}) are required. In this technologically relevant limit, quantum statistics of photons will reveal behavior significantly different from the quasi-classical Poisson behavior exhibited by traditional OAM laser beams. These fundamental as well as technologically relevant problems require the definition of single photon quantum state along with OAM/SAM operators.

In this work, we present an important frontier for quantum structured light involving twisted space-time wavepackets of light. We first construct the wave function of a quantum twisted pulse, as well as a twisted laser beam, from quantum field theory~\cite{loudon2000quantum} instead of from the single-particle Schr\"{o}dinger equation in the first-quantization picture~\cite{bialynicki1994wave,Sipe1995photon,barnett2014optical}. By exploiting the quantum operators of the angular momenta of light~\cite{yang2020quantum}, we evaluate the mean value as well as the quantum uncertainty of the photon spin operator vector.  Apart from the well-established global properties of polarization, we also investigate the quantum properties of the photon spin density vector, i.e., the spin texture, which is a function of space and time. We show that beyond the paraxial approximation, the photon spin density of a Bessel single-photon pulse can exhibit rich spatial texture. Our work builds on previous important work in the field  ~\cite{barnett1994orbital,berry1998paraxial,Monteiro2009angular,li2020spin,Li2009spin,cerjan2011orbital,Holleczek2011classical,bliokh2015transverse,Arnaut2000orbital,Jauregui2005quantum,Calvo2006quantum,Milione2011higher}. Our proposed framework provides a powerful and versatile tool to  engineer the local photon spin and OAM densities of a quantum structured light pulse, specifically for spatiotemporal optical vortices~\cite{Chong2020Generation,Jhajj2016spatiotemporal}.

Non-local spin noise (i.e., spin density correlation) for electrons is a fundamental signature of quantum phases of magnetic condensed matter~\cite{sachdev2007quantum}, specifically in phases of matter such as quantum spin liquids without magnetic order~\cite{Broholm2020quantum}. However, no such quantum spin noise operator has been defined for photons till date. Our theoretical formalism allows us to overcome this challenge. Here, we introduce the quantum correlator of photonic spin density to characterize the nonlocal spin noise in light. This paves the way to explore exotic phases of light with long-range spin order. 

We emphasize that our work is immediately amenable to experimental verification. We predict that for Bessel pulses with large OAM, there will exist large fluctuations in the OAM along orthogonal directions. This additional quantum noise can be verified in metrology experiments even with OAM laser beams. Recently, it was demonstrated that the nitrogen-vacancy (NV) center in diamond can be used as a quantum sensor for detecting the local spinning nature of photons~\cite{Kalhor2020probe}. The spin density of the off-resonant optical beam can induce an effective static magnetic field for the electron spin of the NV center, which itself is an atomic-scale magnetometer working at room temperature. Imaging of our discovered helical spin-density structure in this work can be realized with the same technology in the near future. Furthermore, our proposed non-local spin density correlation can also be measured in compound measurements with two or multiple NV centers.

\section*{Results and Discussion}

\subsection{Quantum Spin and orbital angular momenta of light}
The full quantum operator of photon spin is given by ~\cite{yang2020quantum}
\begin{equation}
\hat{\boldsymbol{S}}=\frac{1}{c}\int d^{3}x\hat{\boldsymbol{\pi}}(\boldsymbol{r},t)\times\hat{\boldsymbol{A}}(\boldsymbol{r},t). \label{eq:SM-lorenz} 
\end{equation}
and OAM $\hat{\boldsymbol{L}}$ of light in the Lorenz gauge within quantum field theory. The operators $\hat{\boldsymbol{S}}$ and $\hat{\boldsymbol{L}}$ obey the canonical commutation relationships
\begin{equation}
[\hat{S}_i,\hat{S}_j]=i\hbar\epsilon_{ijk}\hat{S}_k,\ [\hat{L}_i,\hat{L}_j]=i\hbar\epsilon_{ijk}\hat{L}_k,\ [\hat{L}_i,\hat{S}_j]=0,
\end{equation} 
where $\epsilon_{ijk}$ is the third-order Levi-Civita symbol. The longitudinal and scalar photons play a significant role in both $\hat{\boldsymbol{S}}$ and $\hat{\boldsymbol{L}}$. However, only the SAM $\hat{\boldsymbol{S}}^{\rm obs} =\varepsilon_{0}\int d^{3}r\hat{\boldsymbol{E}}_{\perp}(\boldsymbol{r},t)\times\hat{\boldsymbol{A}}_{\perp}(\boldsymbol{r},t)$ and OAM $\hat{\boldsymbol{L}}^{\rm obs} =\varepsilon_{0}\int d^{3}r\hat{E}_{\perp}^{j}(\boldsymbol{r},t)(\boldsymbol{r}\times\boldsymbol{\nabla})\hat{A}_{\perp}^{j}(\boldsymbol{r},t)$ carried by transversely polarized photons are directly observable quantities even in the presence of charges~\cite{yang2020quantum}.
Note, $\hat{\boldsymbol{E}}_{\perp}$ and $\hat{\boldsymbol{A}}_{\perp}$ are the transverse part of the electric field and the vector potential, respectively.

\begin{figure*}
\centering
\includegraphics[width=16cm]{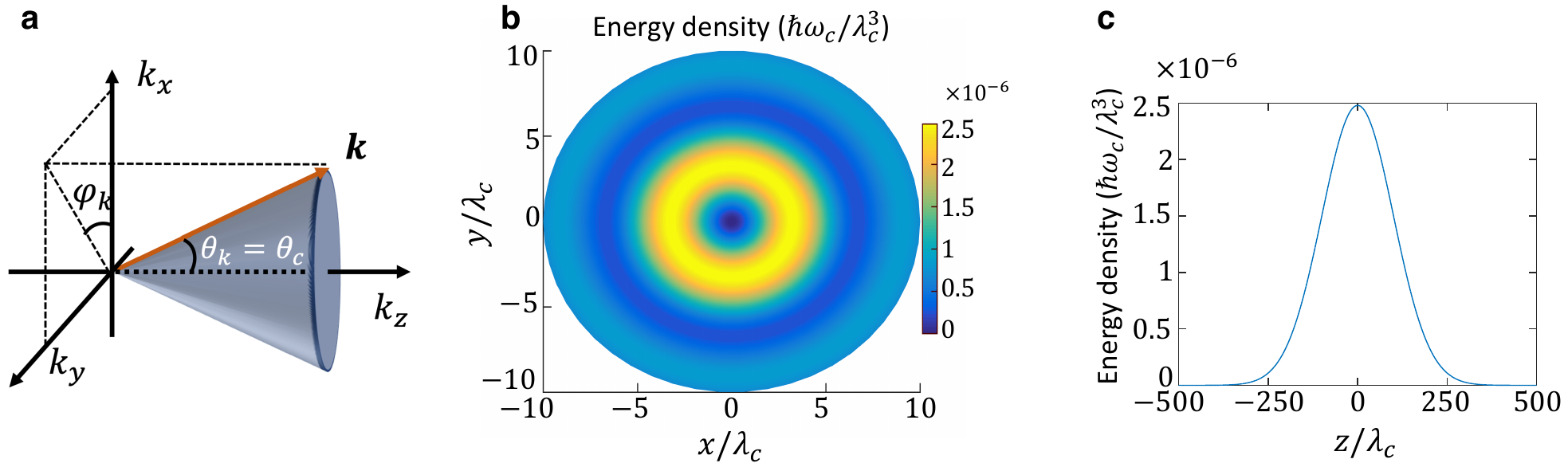}\caption{\label{fig:cone} \textbf{a} \textbf{Profile of a Bessel pulse in wave-vector space and real space.} \textbf{a} Schematic of the spectral distribution of a Bessel pulse. \textbf{b} The energy density (in unit of $\hbar \omega_c/\lambda_c^3$) distribution of a Bessel pulse in the $xy$-plane. \textbf{c} The envelope of the Bessel pulse in the propagating direction is of a Gaussian type. Here, $\omega_c$ is the center frequency of the pulse with center wavelength $\lambda_c$. The polar angle of the Bessel pulse is taken as $\theta_c = 0.2\pi$. The other parameters are taken as $m=2$ and $C_0 = 100$. 
}
\end{figure*}

Using the circularly polarized plane waves, we can expand the observable photon spin and OAM operators as~\cite{cohen1997photons,van1994commutation,van1994spin}
\begin{align}
\hat{\boldsymbol{S}}^{\rm obs} & =\hbar\int d^{3}k\left[\hat{a}_{\boldsymbol{k},+}^{\dagger}\hat{a}_{\boldsymbol{k},+}-\hat{a}_{\boldsymbol{k},-}^{\dagger}\hat{a}_{\boldsymbol{k},-}\right]\boldsymbol{e}(\boldsymbol{k},3),\label{eq:spin}\\
\textcolor{red}{\hat{\boldsymbol{L}}^{\rm obs}} & \textcolor{red}{=-i\hbar\int d^{3}k\sum_{\lambda=\pm}\hat{a}_{\boldsymbol{k},\lambda}^{\dagger}\left[\boldsymbol{k}\times\boldsymbol{D}_{\lambda}\right]\hat{a}_{\boldsymbol{k},\lambda}},
\end{align}
where $\boldsymbol{e}(\boldsymbol{k},3)=\boldsymbol{k}/|\boldsymbol{k}|$ is the unit vector and \textcolor{red}{$\lambda=\pm$ label positive and negative helicity, respectively (see Supplementary Note 1). The covariant derivative is $\boldsymbol{D}_{\lambda}=\boldsymbol{\nabla}_{\boldsymbol{k}}-i\boldsymbol{\mathcal A}_{\lambda}(\boldsymbol{k})$, where $\boldsymbol{\mathcal A}_{\lambda}(\boldsymbol{k})=i\boldsymbol{e}^{*}(\boldsymbol{k},\lambda)\cdot\boldsymbol{\nabla}_{\boldsymbol{k}}\boldsymbol{e}(\boldsymbol{k},\lambda)$ is the polarization-frame connection.}  The ladder operators of the plane wave with wave vector $\boldsymbol{k}$ and polarization $\lambda$ satisfy the bosonic commutation relation $[\hat{a}_{\boldsymbol{k},\lambda},\hat{a}_{\boldsymbol{k}',\lambda'}^{\dagger}]=\delta(\boldsymbol{k}-\boldsymbol{k}')\delta_{\lambda\lambda'}$. The photon helicity is given by
$\hat{\Lambda}=\hbar\int d^{3}k\left[\hat{a}_{\boldsymbol{k},+}^{\dagger}\hat{a}_{\boldsymbol{k},+}-\hat{a}_{\boldsymbol{k},-}^{\dagger}\hat{a}_{\boldsymbol{k},-}\right]$. \textcolor{red}{The observable spin and OAM are separately defined but do not commute. As established by van Enk and Nienhuis~\cite{van1994commutation,van1994spin} (see Supplementary Note 1), they satisfy} \begin{equation}
{\color{red}[\hat{S}^{\rm obs}_i,\hat{L}^{\rm obs}_j] = i\hbar\epsilon_{ijk}\hat{S}^{\rm obs}_k.}
\end{equation}

{\color{red}To show the connection between the angular momentum of photons and electrons}, we define a two-component helicity field operator in real space,
$\hat{\psi}(\boldsymbol{r})=
[\hat{\psi}_{+}(\boldsymbol{r}),\hat{\psi}_{-}(\boldsymbol{r})]^{T}$,
where
\begin{equation}
\hat{\psi}_{\lambda}(\boldsymbol{r})
=\frac{1}{\sqrt{(2\pi)^{3}}}
\int d^{3}k\,
\hat{a}_{\boldsymbol{k},\lambda}
e^{i\boldsymbol{k}\cdot\boldsymbol{r}}.
\end{equation}
For the source-free case, our defined field operator in the Heisenberg picture satisfies the homogeneous wave equation
\begin{equation}
\left(\nabla^{2}-\frac{1}{c^{2}}
\frac{\partial^{2}}{\partial t^{2}}\right)
\hat{\psi}(\boldsymbol{r},t)=0.
\end{equation}
\textcolor{red}{In this two-component helicity representation, the observable photon OAM and SAM can be expressed as}
{\color{red}
\begin{align}
\hat{\boldsymbol L}^{\rm obs}
&=\int d^{3}r\,\hat{\psi}^{\dagger}(\boldsymbol r)
\left[
\boldsymbol r\times\hat{\boldsymbol p}
-\boldsymbol{\mathcal C}(\hat{\boldsymbol p})
\right]\hat{\psi}(\boldsymbol r),
\\
\hat{\boldsymbol S}^{\rm obs}
&=\hbar\int d^{3}r\,\hat{\psi}^{\dagger}(\boldsymbol r)
\hat{\sigma}_z
\frac{\hat{\boldsymbol p}}{|\hat{\boldsymbol p}|}
\hat{\psi}(\boldsymbol r),
\end{align}
}
\textcolor{red}{where $\hat{\boldsymbol p}=-i\hbar\boldsymbol\nabla$ is the momentum operator, $|\hat{\boldsymbol p}|=\sqrt{\hat{\boldsymbol p}^{\,2}}$, and $\hat{\sigma}_z=\mathrm{diag}(1,-1)$ is the Pauli matrix. The polarization-frame correction is defined by}
{\color{red}
\begin{equation}
\boldsymbol{\mathcal C}(\hat{\boldsymbol p})
=
\begin{pmatrix}
\hat{\boldsymbol p}\times
\boldsymbol{\mathcal A}_{+}(\hat{\boldsymbol p}/\hbar)
&0\\
0&
\hat{\boldsymbol p}\times
\boldsymbol{\mathcal A}_{-}(\hat{\boldsymbol p}/\hbar)
\end{pmatrix},
\end{equation}
}\textcolor{red}{Both
$\boldsymbol{\mathcal C}(\hat{\boldsymbol p})$
and $\hat{\boldsymbol p}/|\hat{\boldsymbol p}|$
act as nonlocal Fourier multipliers in real space.
In contrast, the photon helicity admits a simple local expression,}
{\color{red}
\begin{equation}
\hat{\Lambda}
=\hbar\int d^{3}r\,
\hat{\psi}^{\dagger}(\boldsymbol r)
\hat{\sigma}_z
\hat{\psi}(\boldsymbol r).
\end{equation}
}
\textcolor{red}{Consequently, the total observable photon angular momentum reads}
{\color{red}
\begin{align}
\hat{\boldsymbol J}^{\rm obs}
&=\hat{\boldsymbol L}^{\rm obs}
+\hat{\boldsymbol S}^{\rm obs}
\nonumber\\
&=\int d^{3}r\,\hat{\psi}^{\dagger}(\boldsymbol r)
\left[
\boldsymbol r\times\hat{\boldsymbol p}
-\boldsymbol{\mathcal C}(\hat{\boldsymbol p})
+\hbar\hat{\sigma}_z
\frac{\hat{\boldsymbol p}}{|\hat{\boldsymbol p}|}
\right]\hat{\psi}(\boldsymbol r).
\end{align}
}

\subsection{Quantum wave function of twisted light pulses}
In previous sections, we have shown that both $\hat{\boldsymbol{S}}^{\rm obs}$ and $\hat{\boldsymbol{L}}^{\rm obs}$ are vector operators. However, in previous studies, usually only their projections on the propagating direction have been fully studied~\cite{barnett1994orbital,berry1998paraxial,li2020spin}. Their mean value on the transverse plane and more importantly, their quantum fluctuations have not been investigated. On the other hand, the near-field techniques have now been well developed. This makes it possible to measure and engineer the angular-momentum density of light, which is a vector function of space and time, in experiments. Thus, a fully quantum theory beyond the paraxial approximation to explore all classes of twisted pulses in a united framework is highly desirable. Here, we present this powerful theoretical tool by generalizing the quantum theory of continuous-mode field~\cite{loudon2000quantum,Romero2012ultrafast} to the twisted-pulse case. 

We first define the single-photon wave-packet creation operator for a twisted photon pulse 
\begin{equation}
\hat{a}_{\xi\lambda}^{\dagger}=\int d^{3}k\xi_{\lambda}(\boldsymbol{k})\hat{a}_{\boldsymbol{k}\lambda}^{\dagger},
\end{equation}
as a coherent superposition of plane-wave modes. The pulse shape and other quantum properties of the pulse are fully determined by the spectral amplitude function (SAF) $\xi_{\lambda}(\boldsymbol{k})$. In the following, we denote the propagating direction of the pulse as the $z$-axis and work in the cylindrical coordinate in $k$-space $\boldsymbol{k}=k_{z}\boldsymbol{e}_{z}+\rho_{k}\boldsymbol{e}_{\rho}=\rho_{k}\cos\varphi_{k}\boldsymbol{e}_{x}+\rho_{k}\sin\varphi_{k}\boldsymbol{e}_{y}+k_{z}\boldsymbol{e}_{z}$. Here, $\rho_k$ is the radial distance from the $k_z$-axis, $\varphi_k$ is the azimuth angle, and $\boldsymbol{e}$ denotes the corresponding unit vector.  The SAF of a twisted pulse with deterministic OAM can be generally expressed as
\begin{equation}
\xi_{\lambda}(\boldsymbol{k})=\frac{1}{\sqrt{2\pi}}\eta_{\lambda}(k_{z},\rho_{k})e^{im\varphi_{k}}. \label{eq:SAF}
\end{equation}
Usually, the amplitude $\eta_{\lambda}(k_{z},\rho_{k})$ is symmetric in the transverse plane, i.e, it is independent on the azimuth angle~$\varphi_k$. The phase factor $\exp (i m\varphi_k)$ with an integer $m$ will lead to the OAM of light in $z$-direction of a single-photon pulse as shown in the following. 

The SAF is required to satisfy the normalization condition $\int d^{3}k\left|\xi_{\lambda}(\boldsymbol{k})\right|^{2}=1$. This guarantees that $\hat{a}_{\xi\lambda}^{\dagger}$ obeys the bosonic commutation relation
\begin{equation}
[\hat{a}_{\xi\lambda},\hat{a}_{\xi\lambda}^{\dagger}]=1.    
\end{equation}
Then, the wave-packet creation operator $\hat{a}_{\xi\lambda}^{\dagger}$ can be treated as a normal ladder operator of a harmonic oscillator. Using this commutation relation, we can construct the wave function of all classes of quantum pulses in the standard way, such as the most common $n$-photon Fock-state and  coherent-state pulses~\cite{loudon2000quantum} (please refer to Supplementary Note 3)
\begin{equation}
\left|n_{\xi\lambda}\right\rangle =\frac{1}{\sqrt{n!}}\left(\hat{a}_{\xi\lambda}^{\dagger}\right)^{n}\left|0\right\rangle,
\end{equation}
and
\begin{equation}
\left|\alpha_{\xi\lambda}\right\rangle =\exp\left(\alpha\hat{a}_{\xi\lambda}^{\dagger}-\frac{1}{2}\bar{n}\right)\left|0\right\rangle =e^{-\bar{n}/2}\sum_{n=0}^{\infty}\frac{\alpha^{n}}{\sqrt{n!}}\left|n_{\xi\lambda}\right\rangle ,
\end{equation}
where $\bar{n}=|\alpha|^{2}$ is the mean photon number in the coherent-state pulse. The wave function of a squeezed-state pulse, an entangled two-photon pulse~\cite{Torres2003entangled}, or an ultra-short spatiotemporal vortex pulse~\cite{Chong2020Generation,Jhajj2016spatiotemporal} can also be constructed similarly. Here, the polarization of the pulse is fixed as one of the circular polarizations. However, linearly or elliptically polarized quantum pulses can also be constructed with superposition of two circular polarization ladder operators $\hat{a}_{\xi\lambda} \ (\lambda=\pm)$. We also note that a twisted laser beam can be characterized by a wave function with a very long pulse length and very large photon number. Thus, our method also captures the cases of continuous OAM laser beams used widely in experiments.

Without loss of generality, we only take the Bessel pulses as an example to show the quantum properties of the spin and OAM of twisted pulses. Other twisted pulses, such as a Bessel-Gaussian or Laguerre-Gaussian pulse, can be treated similarly.  The single-frequency Bessel beam is the superposition of all plane waves on the cone with the same frequency $\omega=c|\boldsymbol{k}|$, $ k_{z}$, and polar angle $\theta_{k}=\theta_c$ as shown in Fig.~\ref{fig:cone}\textbf{a}. Then, the SAF of a Bessel pulse with a Gaussian envelope can be expressed $\eta_{\lambda}(k_{z},\rho_{k})$
as the product of two Gaussian functions
\begin{align}
\eta_{\lambda}(k_{z},\rho_{k})=&\left(\frac{2\sigma_{z}^{2}}{\pi}\right)^{1/4}\exp\left[-\sigma_{z}^{2}(k_{z}-k_{z,c})^{2}\right] \nonumber \\
&\times\left(\frac{2\sigma_{\rho}^{2}}{\pi k_{\perp,c}^{2}}\right)^{1/4}\exp\left[-\sigma_{\rho}^{2}(\rho_{k}-k_{\perp,c})^{2}\right].
\end{align}
The first Gaussian function with width $1/\sigma_z$ and center wave vector $k_{z,c}$ characterizes the envelope of the pulse in the propagating direction. The pulse length on $z$-axis in real space is given by $\sigma_{z}=c\tau_{p}$  with $\tau_{p}$ the pulse length in time domain (please refer to Supplementary Note 3). We show the energy density of a Bessel pulse in Fig.~\ref{fig:cone}\textbf{b} and \textbf{c}.

Distinct from previous works~\cite{Arnaut2000orbital,Jentschura2011generation}, we do not add a delta function [such as $\delta (\theta_k-\theta_{c})$] in the SAF to characterize its distribution property in the $xy$-plane. This will cause a serious issue that the wave functions of the quantum pulses cannot be normalized, because $\int d^3k|\xi_{\lambda}(\boldsymbol{k})|^2\propto \delta (\theta_k-\theta_{c})$. Instead, we utilize another Gaussian function with width $1/\sigma_{\rho}$ and center value $k_{\perp,c}=k_{z,c}\tan\theta_{c}$. These two Gaussian functions should have the same ratio between center wave-number and the width, i.e. $k_{z,c}\sigma_z= k_{\perp,c}\sigma_{\rho}\equiv C_0$. In the narrow bandwidth limit $C_0\gg 1$, our defined SAF is well normalized (please refer to Supplementary Note 3). We also note that in contrast to the Bessel-mode-based method~\cite{Jauregui2005quantum} which only applies to Bessel beams, our generalized plane-wave-based framework is amenable to unify the theory of all classes of quantum pulses.

\subsection{Quantum statistics of the photon spin}
Traditionally, the angular momentum carried by each photon in a twisted laser beam has been calculated semi-classically via the ratio of angular flux to the energy flux~\cite{barnett1994orbital,berry1998paraxial} and only its projection on the propagating axis has been studied. Although the projection of the photon spin and OAM of a non-paraxial beam on the transverse plane has caused attention recently~\cite{li2020spin,Monteiro2009angular,Li2009spin,cerjan2011orbital,bliokh2015transverse}, a systematic and comprehensive investigation of the vector nature of the photon spin and OAM is still missing. Specifically, the Heisenberg uncertainty relation for photon OAM has never been investigated. On the other hand, many researchers have also tried to establish a quantum theory of the angular momentum of light in the last two decades~\cite{Arnaut2000orbital,Jauregui2005quantum,Calvo2006quantum,Milione2011higher,babiker2018atoms}. However, a fully quantum framework to handle arbitrary quantum pulses beyond the paraxial approximation has not been found. 

We first calculate the mean value of the spin of a Fock-state Bessel pulse with polarization $\lambda$ and photon number $n$ (please refer to Supplementary Note 4),
\begin{align}
\left\langle n_{\xi\lambda}\right|\hat{\boldsymbol{S}}^{{\rm obs}}\left|n_{\xi\lambda}\right\rangle =\hbar n\lambda\left(0,0,\cos\theta_{c}\right).
\end{align}
Here, we see that the magnitude of the spin carried by each circularly polarized photon is usually smaller than $\hbar$ and approaches to $\hbar$ asymptotically in the paraxial limit ( $\theta_c\rightarrow 0$)~\cite{Arnaut2000orbital,Li2009spin}. This is significantly different from the helicity, which is exactly $\hbar$. If the SAF of a pulse is symmetric in the $xy$-plane, then the mean value of the spin in the $xy$-plane vanishes, i.e., $\langle \hat{S}^{\rm obs}_x\rangle = \langle \hat{S}^{\rm obs}_y\rangle=0$. However, we show that the quantum fluctuations of photon spin in the $xy$-plane is not zero. The standard derivations of the spin of an $n$-photon Fock-state Bessel pulse are given by
\begin{align}
\Delta\hat{S}^{\rm obs}_{x}  =\Delta\hat{S}^{\rm obs}_{y}=\hbar\sqrt{n/2}\left|\sin\theta_{c}\right|,\ 
\Delta\hat{S}^{\rm obs}_{z} =0.
\end{align}
This is significantly beyond the previous semi-classical theory~\cite{barnett1994orbital,berry1998paraxial,li2020spin}, in which the quantum statistics of the photon spin cannot be studied.

Similarly, we can evaluate the mean value of the spin of a coherent-state Bessel pulse with polarization $\lambda$ and photon number $\bar{n}=|\alpha|^2$,
\begin{equation}
\left\langle \alpha_{\xi\lambda}\right|\hat{\boldsymbol{S}}^{\rm obs}\left|\alpha_{\xi\lambda}\right\rangle = \hbar\bar{n}\lambda\left(0,0,\cos\theta_{c}\right).
\end{equation}
Here, we see that the average spin carried by each photon is still $\hbar\cos \theta_c$ and the spin's projection on $xy$-plane also vanishes. However, the quantum statistics of the photon spin for a coherent-state pulse is significantly different from that of a Fock-state pulse,
\begin{equation}
\Delta\hat{S}^{\rm obs}_{x} =\Delta\hat{S}^{\rm obs}_{y}=\hbar\sqrt{\bar{n}/2}\left|\sin\theta_{c}\right|,\ 
\Delta\hat{S}^{\rm obs}_{z} =\textcolor{red}{\sqrt{\bar{n}}}\hbar\left|\cos\theta_{c}\right|.
\end{equation}
The Poisson statistics of a coherent pulse leads to non-vanishing $\Delta\hat{S}^{\rm obs}_{z}$ in contrast to a sub-Poisson Fock-state pulse.  

\begin{figure}
\centering
\includegraphics[width=8.5cm]{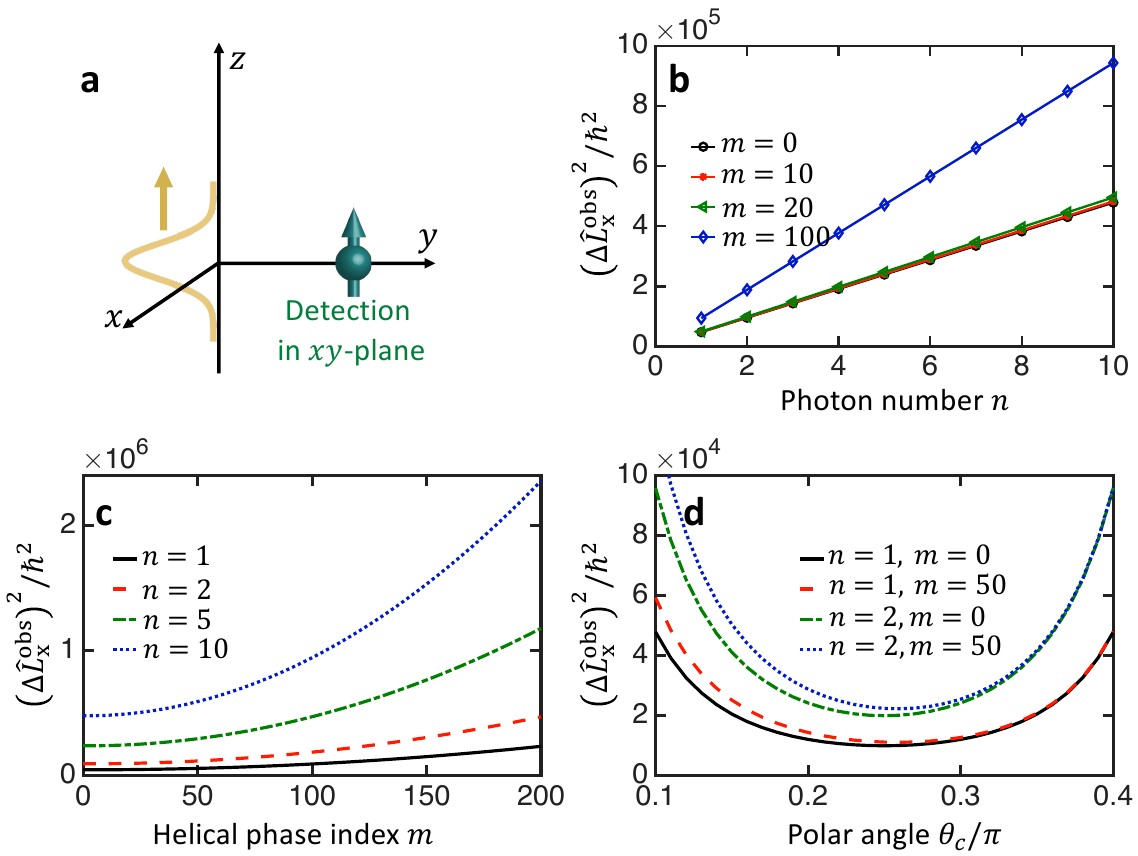}\caption{\label{fig:OAMuncertainty} 
\textbf{Photonic orbital angular momentum (OAM) fluctuation in the transverse plane}.
\textbf{a} OAM measurements in the direction orthogonal to the propagating $z$-axis. A Bessel pulse has vanishing mean OAM in $xy$-plane but non-zero OAM fluctuations, $(\Delta\hat{L}_x^{\rm obs})^2=(\Delta\hat{L}_y^{\rm obs})^2\neq 0$.
\textbf{b} The \textcolor{red}{variance of} $\hat{L}_x^{\rm obs}$ is linearly proportional to the photon number $n$ in the pulse. Different lines correspond to different \textcolor{red}{helical phase indices} $m$.
\textbf{c} The \textcolor{red}{transverse OAM variance exhibits a quadratic dependence on $m-\lambda\cos\theta_c$, reflecting the polarization-frame correction}.
\textbf{d} The \textcolor{red}{transverse OAM variance} is strongly dependent on the polar angle $\theta_c$ of a Bessel pulse. \textcolor{red}{The helicity is fixed at $\lambda=+1$, and} the ratio $C_0$ has been taken as $100$ in all simulations.
}
\end{figure}

\begin{figure*}
\centering
\includegraphics[width=16cm]{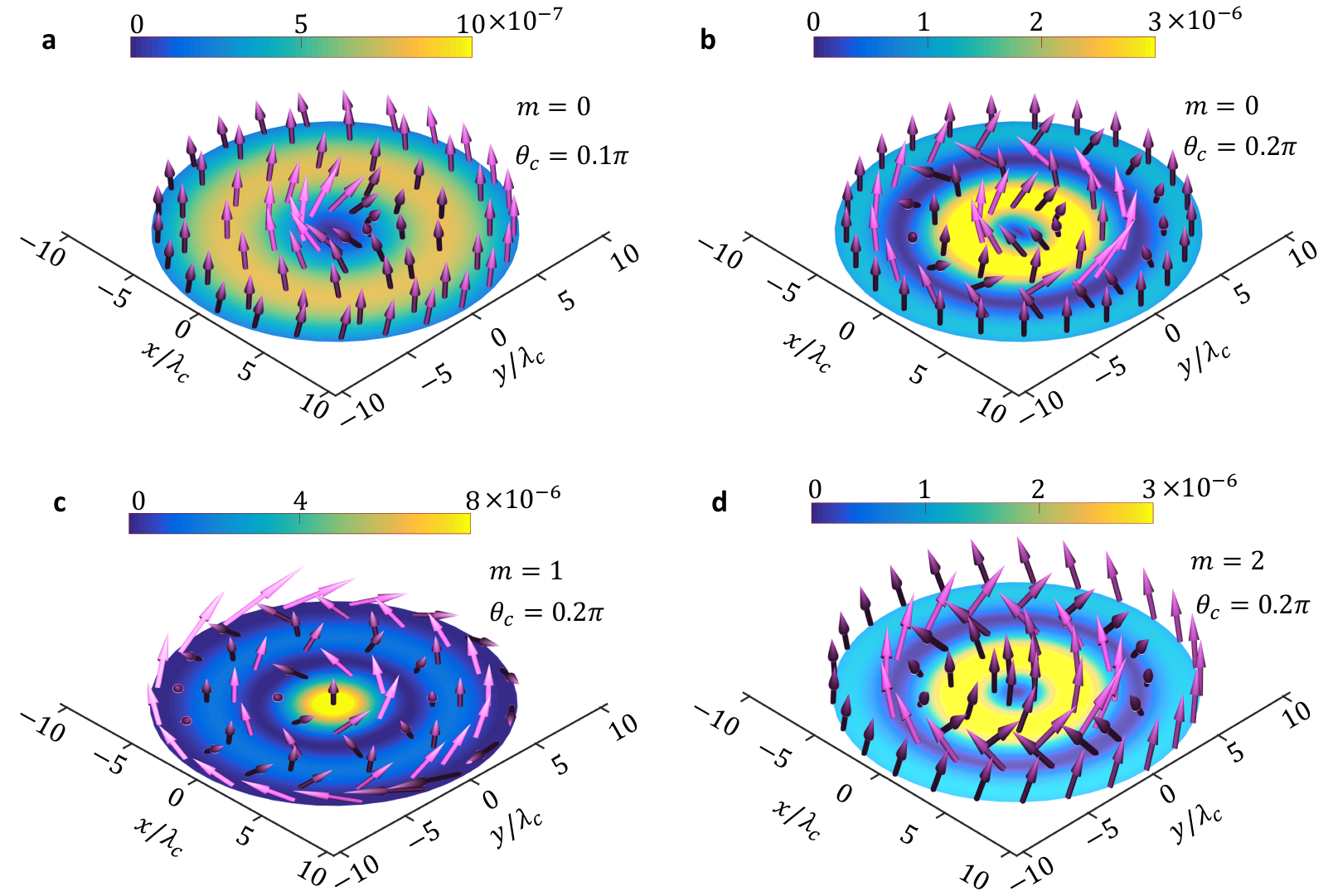}
\caption{\label{fig:spintexture} \textbf{Helical spin texture of twisted quantum pulse.} Spin texture of a single-photon left-circular-polarized Bessel pulse on the pulse-center plane with $k_{z,c}z=ct$. (\textbf{a}-\textbf{d}) correspond to different quantum numbers $m$ and polar angles $\theta_c$. The colorbar describes the amplitude of spin density in unit of $\hbar/\lambda^3_c$.}
\end{figure*}

\subsection{Quantum statistics of the photon OAM}
\textcolor{red}{The observable photon OAM obeys the noncanonical transverse-field spin--orbital algebra, which determines its uncertainty relations.} Here, we present a quantitative investigation about the quantum statistics of photon OAM. \textcolor{red}{For beams with large azimuthal winding numbers, substantial OAM fluctuations can occur in the orthogonal directions, i.e., in the transverse plane.} This quantum effect can be observed in experiment even with traditional OAM laser beams. \textcolor{red}{Using the spherical helicity frame of Supplementary Note 4}, the mean value of $\hat{L}^{\rm obs}_z$ for a Fock-state twisted pulse with photon number $n$ is given by
{\color{red}
\begin{align}
\left\langle n_{\xi\lambda}\right|\hat L_z^{\rm obs}
\left|n_{\xi\lambda}\right\rangle
&=n\int d^3k\,\xi_\lambda^*(\boldsymbol k)\left(-i\hbar\partial_{\varphi_k}
-\hbar\lambda\cos\theta_k\right)\xi_\lambda(\boldsymbol k)\nonumber\\
&=n\hbar(m-\lambda\mu)
\approx n\hbar(m-\lambda\cos\theta_c).
\label{eq:Lz1}
\end{align}
}
\textcolor{red}{Here $\mu\equiv\int d^3k\,|\xi_\lambda(\boldsymbol k)|^2\cos\theta_k$. Thus, the observable longitudinal OAM generally depends on the polarization and the angular spectrum, whereas the total observable angular momentum satisfies $\langle\hat J_z^{\rm obs}\rangle=nm\hbar$ in this helicity frame. The approximation in Eq.~\eqref{eq:Lz1} assumes a narrow cone.} We can also verify that, in this case, the mean value of photon OAM in $xy$-plane vanishes, i.e., $\langle\hat{L}^{\rm obs}_x\rangle=\langle\hat{L}^{\rm obs}_y\rangle=0$ (please refer to Supplementary Note 4).

The quantum variances of the three components of photon OAM for a Fock-state Bessel pulse are given by
{\color{red}
\begin{align}
(\Delta\hat L_z^{\rm obs})^2
\!=\!\langle(\hat L_z^{\rm obs})^2\rangle
 \!-\!\langle\hat L_z^{\rm obs}\rangle^2\!=\!n\hbar^2(\langle\cos^2\theta_k\rangle\!-\!\mu^2)
\approx 0.
\end{align}
}
and
{\color{red}
\begin{align}
&(\Delta\hat L_x^{\rm obs})^2
=(\Delta\hat L_y^{\rm obs})^2\nonumber\\
&\approx\frac{n\hbar^2}{2}\bigg[
\left(C_0^2+\frac14\right)x^2+\frac{C_0^2+\frac34+(m-\lambda\cos\theta_c)^2}{x^2}
-1\bigg],\label{eq:OAMfluctuation}\\
&\ge\frac{n\hbar^2}{2}\bigg[
\sqrt{4C_0^2+1}\sqrt{C_0^2+\frac34}-1\bigg].
\label{eq:inequality}
\end{align}
}where $x=\tan\theta_c\in(0,\infty)$ and \textcolor{red}{$\langle \cos^2\theta_k\rangle\equiv\int d^3k\,|\xi_\lambda(\boldsymbol k)|^2\cos^2\theta_k$. The transverse-variance formula assumes the narrow-band condition $C_0\gg1$ and a smooth spectral cutoff away from the polarization-frame singularity at $\rho_k=0$. The lower bound follows by dropping the nonnegative $(m-\lambda\cos\theta_c)^2/x^2$ term and applying $a^2x^2+b^2/x^2\geq2|ab|$; the angle minimizing the full variance must account for the $\theta_c$ dependence of this term.} \textcolor{red}{Using $[\hat L_x^{\rm obs},\hat L_y^{\rm obs}]=i\hbar(\hat L_z^{\rm obs}-\hat S_z^{\rm obs})$, the appropriate Robertson relation is}
{\color{red}
\begin{align}
\Delta\hat L_x^{\rm obs}\,\Delta\hat L_y^{\rm obs}
&\ge\frac{\hbar}{2}\bigl|\langle\hat L_z^{\rm obs}
-\hat S_z^{\rm obs}\rangle\bigr|=\frac{n\hbar^2}{2}|m-2\lambda\mu|.
\label{eq:uncertainty}
\end{align}
}
The other two \textcolor{red}{Robertson bounds have vanishing right-hand sides for an azimuthally symmetric cone spectrum.} Similar results also hold for a coherent-state twisted pulse, but with non-vanishing \textcolor{red}{$(\Delta\hat{L}^{\rm obs}_{z})^{2}=\hbar^2\bar n\langle(m-\lambda\cos\theta_k)^2\rangle\approx\hbar^2\bar n(m-\lambda\cos\theta_c)^2$.}

Interesting works have been reported to demonstrate the uncertainty relation between the conjugate variables of angle $\varphi$ of light and its derivative $\hat{l}_z\equiv-i\hbar\partial/\partial\varphi$ in the first-quantization picture~\cite{Franke2004Uncertainty,Leach2010quantum}, i.e., $\Delta\varphi_{\theta}\Delta l_z \geq \hbar[1-2\pi P(\theta)]/2$. In contrast, our focus is the \textcolor{red}{three-dimensional uncertainty relations of the observable photon OAM, including its polarization-frame correction.} On the other hand, we note that, for transverse EM fields, \{Hamiltonian $\hat{H}$, momentum $\hat{\boldsymbol{P}}$, helicity $\hat{\Lambda}$\} has been select as as the complete set of commuting observables to specify a photon state usually. \textcolor{red}{A single-photon pulse with fixed azimuthal winding $m$ in the chosen helicity frame is generally not an eigenstate of $\hat L_z^{\rm obs}$ for a finite angular spread, nor of $(\hat{\boldsymbol L}^{\rm obs})^2$.}

Our predicted large OAM fluctuations in $xy$-plane can be verified in experiments (see Fig.~\ref{fig:OAMuncertainty}\textbf{a}). The \textcolor{red}{transverse OAM variances} are linearly proportional to the photon number $n$ in a Bessel pulse as shown in Fig.~\ref{fig:OAMuncertainty}\textbf{b} and \textcolor{red}{have an approximately quadratic dependence on $m-\lambda\cos\theta_c$ at fixed $\theta_c$}, as shown in Fig.~\ref{fig:OAMuncertainty}\textbf{c}. From Eq.~(\ref{eq:OAMfluctuation}), we see that the OAM fluctuations in the transverse plane \textcolor{red}{are} strongly dependent on the polar angle $\theta_c$ of a Bessel pulse. There exists a minimum-uncertainty angle due to \textcolor{red}{the competition between the $x^2$ and $x^{-2}$ contributions in Eq.~(\ref{eq:OAMfluctuation})} as shown in Fig.~\ref{fig:OAMuncertainty}\textbf{d}. For a optical pulse, the ratio $C_0$
between its center wave number and its width is usually very large, e.g., $C_0\approx 188$ for a $50$ fs pulse with center wave length $\lambda_c =500$ nm. In our numerical simulation, we set $C_0=100$. We note that this large OAM fluctuations in the transverse plane also exist in traditional OAM laser beams, such as the routinely used Laguerre-Gaussian beams in experiments.

\begin{figure*}
\centering
\includegraphics[width=18cm]{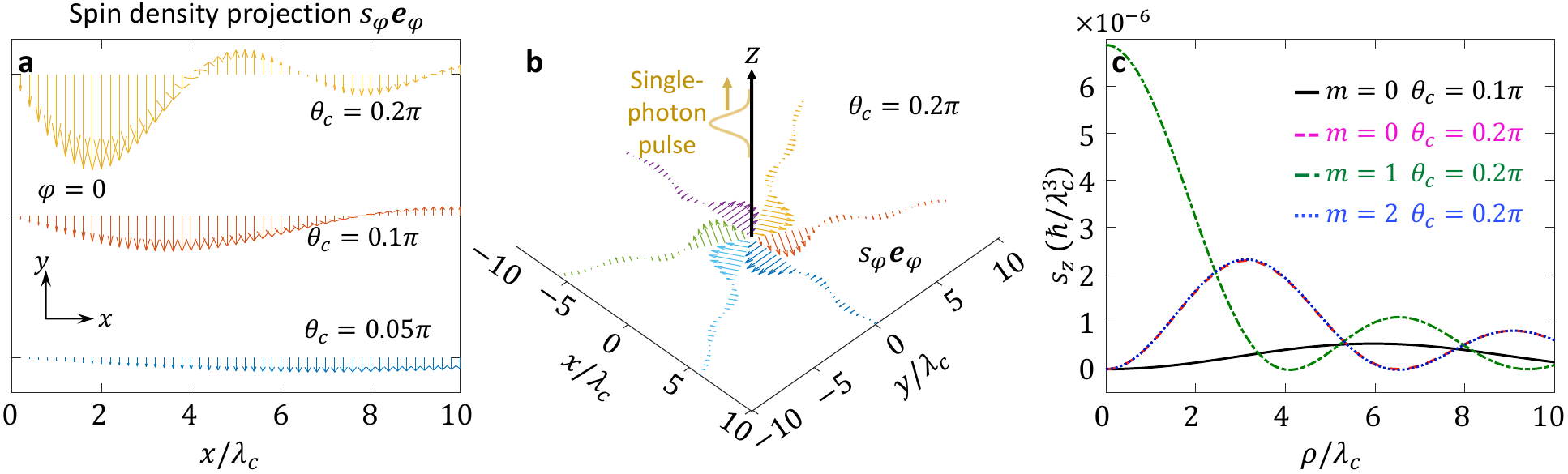}
\caption{\label{fig:projection} \textbf{Details about the spin density vector.} Projections of the spin density of a Bessel single-photon pulse on $xy$-plane (panel \textbf{a} and \textbf{b}) and on $z$-axis in panel \textbf{c}. Panel \textbf{a} corresponds to the case with fixed azimuthal angle $\phi=0$, but different polar angle $\theta_c$. Panel \textbf{b} corresponds to the case with fixed polar angle $\theta_c=0.2\pi$, but different azimuthal angle $\phi = \{0,\pi/3,2\pi/3,\pi,4\pi/3,5\pi/3\}$. Panel \textbf{c} shows the $z$-component of the spin density.}
\end{figure*}

\subsection{Quantum spin texture of a single-photon pulse}
We show that the spin texture of a single-photon pulse can exhibit a very rich and interesting structure in the case beyond the paraxial approximation. The photon spin texture is characterized by the spin density operator
\begin{equation}
\hat{\boldsymbol{s}}^{\rm obs}(\boldsymbol{r},t) =\varepsilon_{0}\hat{\boldsymbol{E}}_{\perp}(\boldsymbol{r},t)\times\hat{\boldsymbol{A}}_{\perp}(\boldsymbol{r},t).    
\end{equation}
Similar to the electric or magnetic fields, the spin density can be treated as a vector field and can be measured locally~\cite{Kalhor2020probe}. We emphasize that as a vector, the spin density is neither purely longitudinal or purely transverse in most cases. In the single-mode plane-wave limit, the spin density will be a space-independent constant, i.e., $\boldsymbol{\nabla}\times\langle\hat{\boldsymbol{s}}^{\rm obs}(\boldsymbol{r},t)\rangle=\boldsymbol{\nabla}\cdot\langle\hat{\boldsymbol{s}}^{\rm obs}(\boldsymbol{r},t)\rangle=0$. 

The mean value of the spin density of a Fock-state Bessel pulse is given (please refer to Supplementary Note 5)
\begin{equation}
\left\langle n_{\xi\lambda}\right|\hat{\boldsymbol{s}}^{\rm obs}(\boldsymbol{r},t)\left|n_{\xi\lambda}\right\rangle  =\lambda\left(s_{\varphi}\boldsymbol{e}_{\varphi}+s_{z}\boldsymbol{e}_{z}\right),\label{eq:helical}    
\end{equation}
where\begin{widetext}
\begin{align}
s_{z} & =\frac{n\hbar C_{0}}{2\pi\sigma_{z}\sigma_{\rho}^{2}}\left\{ \left[J_{m-\lambda}(k_{\perp,c}\rho)\right]^{2}\cos^{4}\frac{\theta_{c}}{2}-\left[J_{m+\lambda}(k_{\perp,c}\rho)\right]^{2}\sin^{4}\frac{\theta_{c}}{2}\right\} \exp\left[-\frac{\left(ct-z\cos\theta_{c}\right)^{2}}{2\sigma_{z}^{2}\cos^{2}\theta_{c}}\right],\label{eq:sz}
\end{align}
and
\begin{align}
s_{\varphi} & =\frac{n\hbar C_{0}\sin\theta_{c}}{2\pi\sigma_{z}\sigma_{\rho}^{2}}\left[\cos^{2}\frac{\theta_{c}}{2}J_{m}(k_{\perp,c}\rho)J_{m-\lambda}(k_{\perp,c}\rho)+\sin^{2}\frac{\theta_{c}}{2}J_{m+\lambda}(k_{\perp,c}\rho)J_{m}(k_{\perp,c}\rho)\right]\exp\left[-\frac{\left(ct-z\cos\theta_{c}\right)^{2}}{2\sigma_{z}^{2}\cos^{2}\theta_{c}}\right],\label{eq:srho}
\end{align}
\end{widetext}
with $\boldsymbol{r}=\rho\boldsymbol{e}_{\rho} + z \boldsymbol{e}_{z}$. The spin density of a coherent pulse can be evaluated similarly. Here, we can see the following key characters of the spin density:
(i) its projection in the $xy$-plane is symmetric around $z$-axis. This causes the corresponding spatial integral to
vanish as shown in previous section, i.e., $\langle\hat{S}^{\rm obs}_{x}\rangle =\langle \hat{S}^{\rm obs}_{y}\rangle =0$;
(ii) its $xy$-plane projection is parallel or anti-parallel to the azimuth-angle-dependent unit vector $\boldsymbol{e}_{\varphi}$ and it does not have a radial component. This leads to the helical spin texture as shown in Fig.~\ref{fig:spintexture}.
(iii) its $xy$-plane projection contains the product of two different Bessel functions. The sign of a Bessel functions flips when crossing its zeros. This leads to the oscillation between clockwise
and anti-clockwise structures in the spin texture; (iv) its projection on $z$ is independent on $\varphi$. For a small
angle $\theta_{c}$, the term $\sim \cos^4(\theta_c /2)$ dominates. Thus, the sign of
$s_{z}$ is always positive (negative) for LCP (RCP) pulse.
This leads to the non-vanishing global spin $\langle \hat{S}_{M,z}\rangle$.

We show the spin texture of an LCP single-photon ($n=1$) Fock-state Bessel pulse in Fig.~\ref{fig:spintexture}. Here, we only look at the spin density vector field on the plane $k_{z,c}z=ct$, at which the Gaussian functions in Eqs.~(\ref{eq:sz}) and (\ref{eq:srho}) reach their maxima. In this case, the space-dependent spin density is only a function of the radius $\rho$ and the azimuthal angle $\varphi$ contained in $\boldsymbol{e}_{\varphi}$. For a pulse with small polar angle $\theta_c=0.1\pi$, almost only a clockwise structure can be observed in panel \textbf{a}. However, for a pulse with larger polar angle~$\theta_c=0.2\pi$, the oscillation between clockwise and counter-clockwise structure can be observed clearly. This oscillation can only be obtained beyond the scalar-field theory and the paraxial approximation. For higher-order Bessel pulses with $m>0$, the fine structure of the spin density is significantly different from the $m=0$ case. The innermost ring changes from clockwise to counter-clockwise as shown in panel \textbf{c} and panel \textbf{d}. We also note that the Bessel pulse with $m=1$ is very special (see panel \textbf{c}), because the spin texture has a peak instead of a hole at the center.

In Fig.~\ref{fig:projection}, we show more details of the projection of the spin density vector field on $xy$-plane and $z$-axis, respectively. In panel \textbf{a}, we look at the projection of the spin density on $xy$-plane $s_{\varphi}\boldsymbol{e}_{
\varphi}$ with fixed Bessel order index as $m=0$ and the azimuthal angle as $\varphi=0$ (i.e., along the $x$-axis). For a pulse with small $\theta_c$ (see the blue arrows at the bottom), $s_{\varphi}\boldsymbol{e}_{
\varphi}$ is relatively small and flat. The amplitude of $s_{\varphi}$ decrease with $\theta_c$ and it vanishes when $\theta_c\rightarrow 0$. For a pulse with larger $\theta_c$ (see the yellow arrows at the top), the sign of $s_{\varphi}$ oscillates between $\pm 1$ with increasing $\rho$. This explains the oscillation between the clockwise and counter-clockwise structure shown in Fig.~{\ref{fig:spintexture}} \textbf{b}. In panel \textbf{b}, we show the rotation of $s_{\varphi}\boldsymbol{e}_{
\varphi}$ in $xy$-plane with fixed $m=0$ and $\theta_c =0.2\pi$. In panel \textbf{c}, we show the projection of the spin density on $z$-axis as the function of $\rho$ for the four cases in Fig.~\ref{fig:spintexture}. Here, we clearly see the oscillation induced by the Bessel function in Eq.~(\ref{eq:sz}). Specifically, the vertex at the center for $m=1$.

\begin{figure}
\centering
\includegraphics[width=8.5cm]{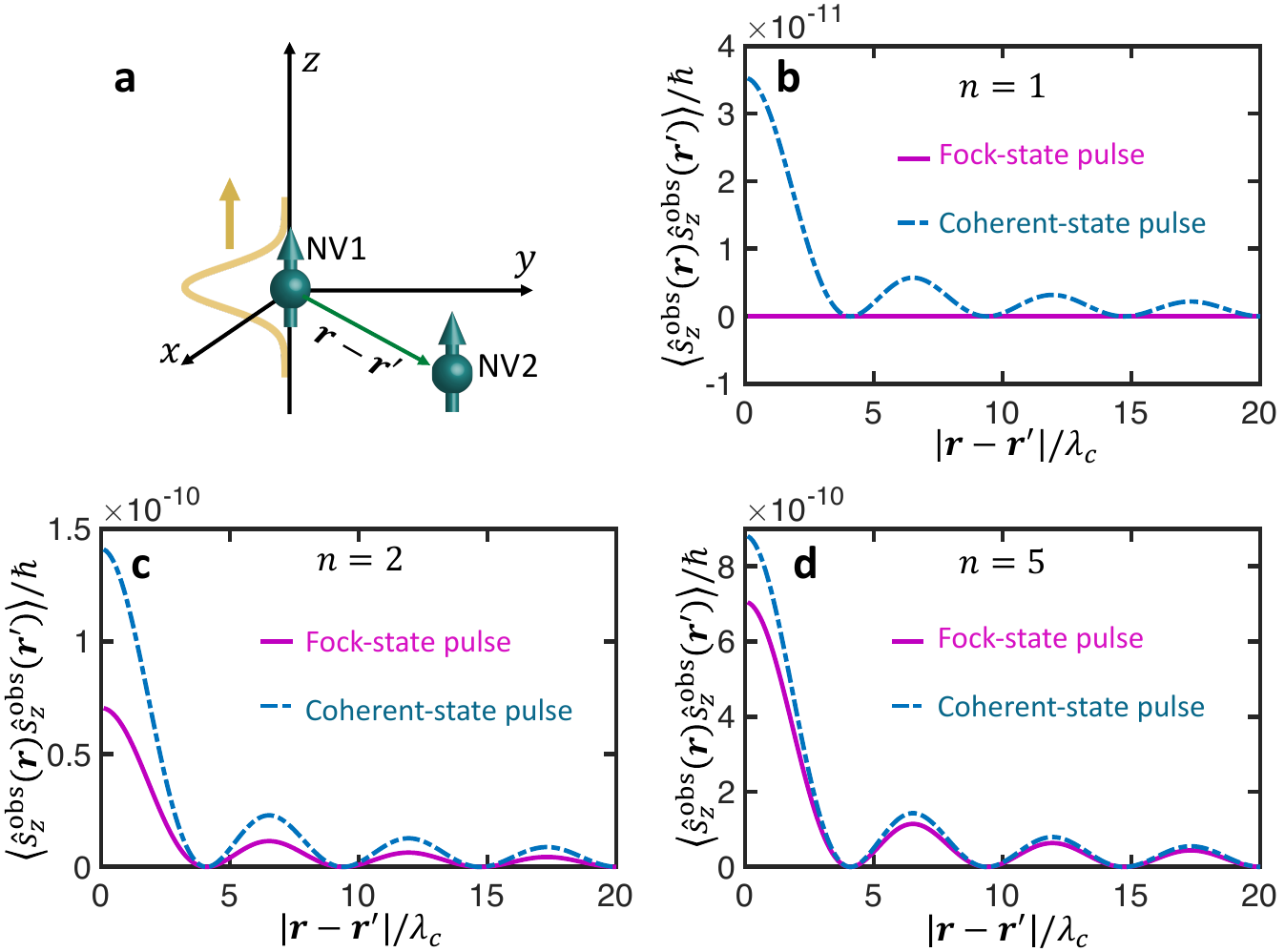}\caption{\label{fig:Correlation} \textbf{Non-local spin noise.} \textbf{Non-local spin noise.} \textbf{a} Our proposed spin density correlation measurement with two nitrogen-vacancy (NV) centers (NV1 and NV2) at $\boldsymbol{r}$ and $\boldsymbol{r}'$, respectively. \textbf{b-d} Contrast of spin density correlation between Fock-state and coherent-state pulses. Here, $n$ is the mean photon number in each pulse and we show the results for the case with $m=1$ and $\theta_c = 0.2\pi$. One of the quantum sensor is fixed on the $z$-axis and the other sensor can move in $xy$-plane. }
\end{figure}

\subsection{Nonlocal spin noise of light}
To characterize the nonlocal spin noise of light, we introduce the quantum correlation function of the photon spin density. Due to the vector nature of the spin density, the full two-point correlation should be characterized by a $3\times 3$ correlation matrix as shown in the Supplementary Note 6. Here, we only describe the equal-time correlator $\left\langle \hat{s}^{\rm obs}_{z}(\boldsymbol{r},t)\hat{s}^{\rm obs}_{z}(\boldsymbol{r}',t)\right\rangle$. 

In the paraxial limit ($\theta_c\approx 0$), the two-point correlation functions for a Fock-state and a coherent-state pulse are given by (please refer to Supplementary Note 6)
\begin{align}
 \left\langle n_{\xi\lambda}\right|\hat{s}^{\rm obs}_{z}(\boldsymbol{r},t)&\hat{s}^{\rm obs}_{z}(\boldsymbol{r}',t)\left|n_{\xi\lambda}\right\rangle \approx \hbar^{2}\left[ \delta(\boldsymbol{r}-\boldsymbol{r}')n\left|\psi_{-\lambda}(\boldsymbol{r},t)\right|^{2}\right. \nonumber \\
  & \left. \ \ \ + n(n-1)\left|\psi_{-\lambda}(\boldsymbol{r},t)\right|^{2}\left|\psi_{-\lambda}(\boldsymbol{r}',t)\right|^{2}\right],
\end{align}
and
\begin{align}
\left\langle \alpha_{\xi\lambda}\right|\hat{s}^{\rm obs}_{z}(\boldsymbol{r},t)\hat{s}^{\rm obs}_{z}&(\boldsymbol{r}',t)\left|\alpha_{\xi\lambda}\right\rangle 
\approx  \hbar^{2}\left[\delta(\boldsymbol{r}-\boldsymbol{r}')\bar{n}\left|\psi_{-\lambda}(\boldsymbol{r},t)\right|^{2}\right. \nonumber \\
& \left. + \bar{n}^{2}\left|\psi_{-\lambda}(\boldsymbol{r},t)\right|^{2}\left|\psi_{-\lambda}(\boldsymbol{r}',t)\right|^{2}\right],
\end{align}
where 
\begin{equation}
\psi_{\lambda}(\boldsymbol{r},t)= \frac{1}{\sqrt{(2\pi)^{3}}}\int d^{3}k\xi_{\lambda}(\boldsymbol{k})e^{i[(\boldsymbol{k}\cdot\boldsymbol{r}-\omega t)+\lambda\varphi_{k}]},
\end{equation}
is the effective wave function of a pulse in real space. This method can be easily generalized to higher-order correlations.

We note that the delta function $\delta(\boldsymbol{r}-\boldsymbol{r}')$ in the correlation function will not lead to any diverging effect, because a practical probe always measures the averaged photon spin density over a finite volume instead of the true single-point spin density. On the other hand, this term vanishes in a composite measurement with $\boldsymbol{r}\neq \boldsymbol{r}'$. In this case, we see that the Poisson and sub-Poisson statistics automatically enter the quantum spin-density correlations. Specifically, the two-point spin density correlation vanishes for a single-photon Fock-state pulse as expected.

We now propose to detect the non-local spin density correlation via compound measurements of two NV centers, which have been exploited as nano-scale quantum sensors for photonic spin density measurements  recently~\cite{Kalhor2020probe}. As shown in Fig.~\ref{fig:Correlation}\textbf{a}, we fixed one quantum sensor on the $z$-axis and move the other one to image the distribution of the spin density correlation in the transverse plane. We contrast the spin density correlations in Fock-state and coherent-state Bessel pules in Fig.~\ref{fig:Correlation}\textbf{b-d}. Here, we see that in the few-photon limit, there exist significant differences between Fock-state and coherent pulses. This difference fundamentally roots in the quantum statistics of photons and it will disappear in the large-photon limit.

The electronic ground-state of a negatively charged NV center is a spin-1 system, which has been routinely used as a highly sensitive nano-scale magnetometer at room temperature~\cite{barry2016optical}. A laser with wavelength shorter than 637 nm is required to excite the NV to its electronic excited states. A red circularly polarized laser pulse (target pulse) with wavelength around 800 nm will not excite the NV, but only induce energy shifts in the three ground spin states.  Recent work has shown that these energy shifts function as a static magnetic field for the NV spin~\cite{Kalhor2020probe}, which is linearly proportional to the local spin density of the target beam, i.e., $\boldsymbol{B}_{\rm eff}\propto \langle\hat{\boldsymbol{s}}^{\rm obs}(\boldsymbol{r})\rangle$. Thus, an NV center can be exploited as a nano-scale photonic spin sensor.

Currently, imaging of single-photon level spin density and the corresponding correlation is extremely challenging in experiments. However, our discovered interesting texture of spin density and non-local spin noise also exists in traditional OAM beam, which can be demonstrated in the near future. On the other hand, due to the absence of photon-photon interaction, the nonlocal spin noise within a light pulse in free space is fully determined by the photon-number statistics.
However, we predict that exotic photonic phases with long-range spin order can exist in a quantum polariton system or an atomic lattice~\cite{chang2017exponential,perczel2017topological}.

\textbf{Conclusion}

We have established the fully quantum framework for photonic angular momenta of quantum structured pulses, as well as the corresponding quantum texture. Our approach presents a paradigm shift for the photonics community as it can be exploited to study the quantum properties and to reveal the vector nature of angular momentum of light. We have shown that the spin texture of a Bessel pulse can exhibit very interesting structure beyond the paraxial limit. Our proposed non-local spin noise will open a new frontier for studying exotic phases of photons with long-range spin order. This spin noise can be measured in compound measurements with multiple nano-scale spin sensors, which have been proposed and demonstrated in our previous experiment~\cite{Kalhor2020probe}. The photonic OAM density and the corresponding non-local OAM density noise can also be handled within our proposed theoretical framework, which will be addressed in our future work.

{\bf ACKNOWLEDGEMENTS}\\
This work is supported by the funding from DARPA Nascent Light-Matter Interactions. L.P.Y is also supported by the funding from Ministry of Science and Technology of China (No. 2021YFE0193500).

\bibliography{main}


\widetext
\newpage
\begin{center}
\textbf{\large Supplementary Materials for ``Quantum structured light: Non-classical spin texture of twisted single-photon pulses''}
\end{center}

\maketitle

\tableofcontents

\noindent\textbf{Supplementary Note 1: Quantum operator of photon angular momentum and helicity}
\vspace{0.5cm}

In this section, we give the quantum operators of the photon angular momentum. First, we give the plane-wave expansion of the electromagnetic (EM) field operators
in free space
\begin{align}
\hat{\boldsymbol{A}}_{\perp}(\boldsymbol{r},t) & =\int d^{3}k\sum_{\lambda=1}^{2}\sqrt{\frac{\hbar}{2\varepsilon_{0}\omega(2\pi)^{3}}}\left[\hat{a}_{\boldsymbol{k},\lambda}\boldsymbol{e}(\boldsymbol{k},\lambda)e^{i(\boldsymbol{k}\cdot\boldsymbol{r}-\omega t)}+{\rm h.c.}\right],\\
\hat{\boldsymbol{E}}_{\perp}(\boldsymbol{r},t) & =i\int d^{3}k\sum_{\lambda=1}^{2}\sqrt{\frac{\hbar\omega}{2\varepsilon_{0}(2\pi)^{3}}}\left[\hat{a}_{\boldsymbol{k},\lambda}\boldsymbol{e}(\boldsymbol{k},\lambda)e^{i(\boldsymbol{k}\cdot\boldsymbol{r}-\omega t)}-{\rm h.c.}\right],\\
\hat{\boldsymbol{B}}(\boldsymbol{r},t) & =\frac{i}{c}\int d^{3}k\sqrt{\frac{\hbar\omega}{2\varepsilon_{0}(2\pi)^{3}}}\left\{ \left[\hat{a}_{\boldsymbol{k},1}\boldsymbol{e}(\boldsymbol{k},2)-\hat{a}_{\boldsymbol{k},2}\boldsymbol{e}(\boldsymbol{k},1)\right]e^{i(\boldsymbol{k}\cdot\boldsymbol{r}-\omega t)}-{\rm h.c.}\right\} ,
\end{align}
where $\varepsilon_0$ is the vacuum permittivity, $\omega=c|\boldsymbol{k}|$ is the frequency of the mode with wave vector $\boldsymbol{k}$ and $\boldsymbol{e}(\boldsymbol{k},\lambda)$ ($\lambda = 1,2$) are the unit polarization vectors of the two transverse modes for each $\boldsymbol{k}$, i.e., $\boldsymbol{e}(\boldsymbol{k},1)\cdot\boldsymbol{k}=\boldsymbol{e}(\boldsymbol{k},2)\cdot\boldsymbol{k}=\boldsymbol{e}(\boldsymbol{k},1)\cdot\boldsymbol{e}(\boldsymbol{k},2)=0$. \textcolor{red}{We choose a locally defined real, right-handed polarization frame, $\boldsymbol e(\boldsymbol k,1)\times\boldsymbol e(\boldsymbol k,2)=\hat{\boldsymbol k}$.} Here, we \textcolor{red}{work} in the Heisenberg picture, thus all field operators are time-dependent.

Using the bosonic commutation relations for the ladder operators
\begin{equation}
    [\hat{a}_{\boldsymbol{k},\lambda},\hat{a}_{\boldsymbol{k}',\lambda'}^{\dagger}]=\delta(\boldsymbol{k}-\boldsymbol{k}')\delta_{\lambda,\lambda'},\ [\hat{a}_{\boldsymbol{k},\lambda},\hat{a}_{\boldsymbol{k}',\lambda'}] = [\hat{a}_{\boldsymbol{k},\lambda}^{\dagger},\hat{a}_{\boldsymbol{k}',\lambda'}^{\dagger}] =0,
\end{equation}
we can verify the following equal-time commutation relations
\begin{equation}
[\hat{A}_{\perp}^i(\boldsymbol{r},t),\hat{E}^j_{\perp}(\boldsymbol{r}',t)] = -\frac{i\hbar}{\varepsilon_0} \textcolor{red}{\delta^{\perp}_{ij}(\boldsymbol{r}-\boldsymbol{r}')} ,\ [\hat{A}_{\perp}^i(\boldsymbol{r},t),\hat{A}^j_{\perp}(\boldsymbol{r}',t)] = [\hat{E}_{\perp}^i(\boldsymbol{r},t),\hat{E}^j_{\perp}(\boldsymbol{r}',t)] =0,
\end{equation}
where we have used the identity \textcolor{red}{$\sum_{\lambda=1}^{2}e_i(\boldsymbol k,\lambda)e_j(\boldsymbol k,\lambda)=\delta_{ij}-k_i k_j/|\boldsymbol k|^2$}. \textcolor{red}{The transverse delta distribution is defined by}
{\color{red}
\begin{equation*}
\delta^\perp_{ij}(\boldsymbol R)=\int \frac{d^3k}{(2\pi)^3}
\left(\delta_{ij}-\frac{k_i k_j}{|\boldsymbol k|^2}\right)e^{i\boldsymbol k\cdot\boldsymbol R}.
\end{equation*}
}

In our previous work~\cite{yang2020quantum}, we discovered the quantum operator of the full photon spin,
\begin{equation}
\hat{\boldsymbol{S}}=\frac{1}{c}\int d^{3}x\hat{\boldsymbol{\pi}}(\boldsymbol{r},t)\times\hat{\boldsymbol{A}}(\boldsymbol{r},t). \label{eq:SM-lorenz} 
\end{equation}
\textcolor{red}{Here, $\hat{\boldsymbol\pi}$ and $\hat{\boldsymbol A}$ denote the unprojected Lorenz-gauge field operators.}
We have shown that the well-known spin and orbital angular momentum (OAM) of light
\begin{align}
\hat{\boldsymbol{S}}^{\rm obs} & =\varepsilon_{0}\int d^{3}r\hat{\boldsymbol{E}}_{\perp}(\boldsymbol{r},t)\times\hat{\boldsymbol{A}}_{\perp}(\boldsymbol{r},t)=i\hbar\int d^{3}k\left[\hat{a}_{\boldsymbol{k},2}^{\dagger}\hat{a}_{\boldsymbol{k},1}-\hat{a}_{\boldsymbol{k},1}^{\dagger}\hat{a}_{\boldsymbol{k},2}\right]\boldsymbol{e}(\boldsymbol{k},3),\\
\hat{\boldsymbol{L}}^{\rm obs} & =\varepsilon_{0}\int d^{3}r\hat{E}_{\perp}^{j}(\boldsymbol{r},t)(\boldsymbol{r}\times\boldsymbol{\nabla})\hat{A}_{\perp}^{j}(\boldsymbol{r},t)\nonumber\\
&\textcolor{red}{=-i\hbar\int d^{3}k\sum_{\lambda,\lambda'=1}^{2}\hat{a}_{\boldsymbol{k},\lambda}^{\dagger}\left[\delta_{\lambda\lambda'}(\boldsymbol{k}\times\boldsymbol{\nabla}_{\boldsymbol{k}})+\boldsymbol{\Gamma}_{\lambda\lambda'}^{\perp}(\boldsymbol{k})\right]\hat{a}_{\boldsymbol{k},\lambda'}},
\end{align}
are the directly observable part of photon angular momentum. \textcolor{red}{The polarization-frame connection $\boldsymbol\Gamma^{\perp}_{\lambda\lambda'}(\boldsymbol k)=\sum_{j=1}^{3}e_j(\boldsymbol k,\lambda)(\boldsymbol k\times\boldsymbol\nabla_{\boldsymbol k})e_j(\boldsymbol k,\lambda')$ need not vanish when the basis depends on $\boldsymbol k$.} Here,  $\boldsymbol{e}(\boldsymbol{k},3)=\boldsymbol{k}/|\boldsymbol{k}|$ is the
unit vector in the propagating direction of each mode. In this paper, we will only study the observable part of photon angular momentum $\hat{\boldsymbol{S}}^{\rm obs}$ and $\hat{\boldsymbol{L}}^{\rm obs}$, which both are gauge invariant. The helicity of the photon
is defined as
\begin{equation}
\hat{\Lambda}=i\hbar\int d^{3}k\left[\hat{a}_{\boldsymbol{k},2}^{\dagger}\hat{a}_{\boldsymbol{k},1}-\hat{a}_{\boldsymbol{k},1}^{\dagger}\hat{a}_{\boldsymbol{k},2}\right],
\end{equation}
which is a Lorentz invariant scalar \textcolor{red}{under proper orthochronous Lorentz transformations (and is odd under parity)}.

We emphasize that there exist significant differences between the spin and OAM of photons. The photon spin angular momentum fundamentally originates from the rotation between different polarization modes, which generates the relative phase between the plane-wave modes with the same momentum $\boldsymbol{k}$. \textcolor{red}{By contrast, OAM describes spatial-mode rotations in momentum space and includes the polarization-frame connection when the basis depends on $\boldsymbol{k}$. The operators $\hat{\boldsymbol{S}}^{\rm obs}$ and $\hat{\boldsymbol{L}}^{\rm obs}$ characterize the observable spin and orbital contributions to the total angular momentum, respectively, although their Cartesian components do not, in general, form a mutually commuting set. Physical photons carry intrinsic helicity rather than an independent three-component rest-frame spin.}

\subsection*{Commutation relations}
Now, we check the commutation relations between the angular momenta of light. This can be done much more easily in the k-space. We first check the photon spin
\begin{align}
[\hat{S}^{\rm obs}_{i},\hat{S}^{\rm obs}_{j}] & =-\hbar^{2}\int d^{3}k\int d^{3}k'\left[\hat{a}_{\boldsymbol{k},2}^{\dagger}\hat{a}_{\boldsymbol{k},1}-\hat{a}_{\boldsymbol{k},1}^{\dagger}\hat{a}_{\boldsymbol{k},2},\hat{a}_{\boldsymbol{k}',2}^{\dagger}\hat{a}_{\boldsymbol{k}',1}-\hat{a}_{\boldsymbol{k}',1}^{\dagger}\hat{a}_{\boldsymbol{k}',2}\right]e_{i}(\boldsymbol{k},3)e_{j}(\boldsymbol{k}',3)\\
 & =\hbar^{2}\int d^{3}k\int d^{3}k'\left(\hat{a}_{\boldsymbol{k},2}^{\dagger}\hat{a}_{\boldsymbol{k},2}-\hat{a}_{\boldsymbol{k},1}^{\dagger}\hat{a}_{\boldsymbol{k},1}+\hat{a}_{\boldsymbol{k},1}^{\dagger}\hat{a}_{\boldsymbol{k},1}-\textcolor{red}{\hat{a}_{\boldsymbol{k},2}^{\dagger}\hat{a}_{\boldsymbol{k},2}}\right)\delta(\boldsymbol{k}-\boldsymbol{k}')e_{i}(\boldsymbol{k},3)e_{j}(\boldsymbol{k}',3)\\
 & =0.
\end{align}
This recovers the result obtained by van Enk and Nienhuis in 1994~\cite{van1994commutation,van1994spin}. For the photon OAM operator, \textcolor{red}{the momentum dependence of the transverse polarization frame must be retained. Using the helicity modes $h=\pm1$ defined below, introduce the polarization connection and covariant derivative}
{\color{red}
\begin{equation}
\boldsymbol{\mathcal A}_{h}(\boldsymbol{k})
=i\boldsymbol e^{*}(\boldsymbol{k},h)\cdot\boldsymbol\nabla_{\boldsymbol k}
\boldsymbol e(\boldsymbol{k},h),\qquad
\boldsymbol D_{h}=\boldsymbol\nabla_{\boldsymbol k}-i\boldsymbol{\mathcal A}_{h},\qquad
\boldsymbol\ell_{h}=-i\hbar\boldsymbol k\times\boldsymbol D_h.
\end{equation}
}
\textcolor{red}{For the local right-handed polarization frame, the helicity Berry curvature is $\boldsymbol\nabla_{\boldsymbol k}\times\boldsymbol{\mathcal A}_h=-h\hat{\boldsymbol k}/|\boldsymbol k|^2$. Consequently, the one-photon orbital generators satisfy}
{\color{red}
\begin{equation}
[\ell_{h,i},\ell_{h,j}]
=i\hbar\epsilon_{ijk}(\ell_{h,k}-\hbar h\hat{k}_k).
\end{equation}
\begin{equation}
[\hat L^{\rm obs}_{i},\hat L^{\rm obs}_{j}]
=i\hbar\epsilon_{ijk}(\hat L^{\rm obs}_{k}-\hat S^{\rm obs}_{k}).
\end{equation}
}
\textcolor{red}{Likewise, since $[\ell_{h,i},\hbar h\hat{k}_j]
=i\hbar\epsilon_{ijk}\hbar h\hat{k}_k$, the spin--OAM cross commutator is}
{\color{red}
\begin{equation}
[\hat S^{\rm obs}_{i},\hat L^{\rm obs}_{j}]
=i\hbar\epsilon_{ijk}\hat S^{\rm obs}_{k}.
\end{equation}
}
\textcolor{red}{Thus, the observable spin and OAM do not separately generate the standard $SO(3)$ algebra, whereas their sum $\hat{\boldsymbol J}^{\rm obs}=\hat{\boldsymbol L}^{\rm obs}+\hat{\boldsymbol S}^{\rm obs}$ does. These results agree with van Enk and Nienhuis~\cite{van1994commutation,van1994spin}.}

\subsection*{Circular polarization representation}
The main task of this work is to explore the quantum properties of the photon spin. Thus, it will much more convenient to use circular-polarization basis. Now, we define two circularly polarized modes for a given wave vector $\boldsymbol{k}$
\begin{align}
\hat{a}_{\boldsymbol{k},+} & =\frac{1}{\sqrt{2}}\left(\hat{a}_{\boldsymbol{k},1}-i\hat{a}_{\boldsymbol{k},2}\right),\\
\hat{a}_{\boldsymbol{k},-} & =\frac{1}{\sqrt{2}}\left(\hat{a}_{\boldsymbol{k},1}+i\hat{a}_{\boldsymbol{k},2}\right),
\end{align}
 with corresponding polarization vectors
\begin{align}
\boldsymbol{e}(\boldsymbol{k},+) & =\frac{1}{\sqrt{2}}\left[\boldsymbol{e}(\boldsymbol{k},1)+i\boldsymbol{e}(\boldsymbol{k},2)\right],\\
\boldsymbol{e}(\boldsymbol{k},-) & =\frac{1}{\sqrt{2}}\left[\boldsymbol{e}(\boldsymbol{k},1)-i\boldsymbol{e}(\boldsymbol{k},2)\right].
\end{align}
\textcolor{red}{With the chosen right-handed polarization frame, $\boldsymbol{k}\times\boldsymbol e(\boldsymbol{k},h)=-ih|\boldsymbol k|\boldsymbol e(\boldsymbol{k},h)$, so $h=+1$ and $h=-1$ denote positive and negative helicity, respectively.} 

The corresponding EM field operators are given by
\begin{align}
\hat{\boldsymbol{A}}_{\perp}(\boldsymbol{r},t) & =\int d^{3}k\sum_{\lambda=\pm}\sqrt{\frac{\hbar}{2\varepsilon_{0}\omega(2\pi)^{3}}}\left[\hat{a}_{\boldsymbol{k},\lambda}\boldsymbol{e}(\boldsymbol{k},\lambda)e^{i(\boldsymbol{k}\cdot\boldsymbol{r}-\omega t)}+{\rm h.c.}\right].\\
\hat{\boldsymbol{E}}_{\perp}(\boldsymbol{r},t) & =i\int d^{3}k\sum_{\lambda=\pm}\sqrt{\frac{\hbar\omega}{2\varepsilon_{0}(2\pi)^{3}}}\left[\hat{a}_{\boldsymbol{k},\lambda}\boldsymbol{e}(\boldsymbol{k},\lambda)e^{i(\boldsymbol{k}\cdot\boldsymbol{r}-\omega t)}-{\rm h.c.}\right],\\
\hat{\boldsymbol{B}}(\boldsymbol{r},t) & =\frac{1}{c}\int d^{3}k\sqrt{\frac{\hbar\omega}{2\varepsilon_{0}(2\pi)^{3}}}\left\{ \left[\hat{a}_{\boldsymbol{k},+}\boldsymbol{e}(\boldsymbol{k},+)-\hat{a}_{\boldsymbol{k},-}\boldsymbol{e}(\boldsymbol{k},-)\right]e^{i(\boldsymbol{k}\cdot\boldsymbol{r}-\omega t)}+{\rm h.c.}\right\} .
\end{align}

The photon spin, OAM, and helicity can be re-expressed as
\begin{align}
\hat{\boldsymbol{S}}^{\rm obs} & =\hbar\int d^{3}k\left[\hat{a}_{\boldsymbol{k},+}^{\dagger}\hat{a}_{\boldsymbol{k},+}-\hat{a}_{\boldsymbol{k},-}^{\dagger}\hat{a}_{\boldsymbol{k},-}\right]\boldsymbol{e}(\boldsymbol{k},3),\label{eq:spin}\\
\hat{\boldsymbol{L}}^{\rm obs} & =-i\hbar\int d^{3}k\sum_{\lambda=\pm}\hat{a}_{\boldsymbol{k},\lambda}^{\dagger}{\color{red}[\boldsymbol{k}\times\boldsymbol D_{\lambda}]}\hat{a}_{\boldsymbol{k},\lambda},\\
\hat{\Lambda}_{M} & =\hbar\int d^{3}k\left[\hat{a}_{\boldsymbol{k},+}^{\dagger}\hat{a}_{\boldsymbol{k},+}-\hat{a}_{\boldsymbol{k},-}^{\dagger}\hat{a}_{\boldsymbol{k},-}\right]
\end{align}

For convenience, we will mainly work in the cylindrical coordinate frame in $k$-space
\begin{equation}
\boldsymbol{k}=k_{z}\boldsymbol{e}_{z}+\rho_{k}\boldsymbol{e}_{\rho}=\rho_{k}\cos\varphi_{k}\boldsymbol{e}_{x}+\rho_{k}\sin\varphi_{k}\boldsymbol{e}_{y}+k_{z}\boldsymbol{e}_{z},
\end{equation}
where $\rho_k$ is the radial distance from the $k_z$-axis, $\varphi_k$ is the azimuth angle, and $\boldsymbol{e}$ denotes the corresponding unit vector.

\vspace{0.5cm}
\noindent\textbf{Supplementary Note 2: Photon OAM operator in reciprocal space}
\vspace{0.5cm}

In this section, we give the quantum operators of photon OAM in reciprocal space. \textcolor{red}{For a momentum-dependent transverse polarization frame, the three components must include the polarization-frame connection.} They can be written as
{\color{red}
\begin{align}
\hat L^{\rm obs}_{\alpha}
&=\int d^3k\sum_{r,s=1}^{2}
\hat a_{\boldsymbol k,r}^{\dagger}
\hat{\mathcal L}_{\alpha,rs}(\boldsymbol k)
\hat a_{\boldsymbol k,s},\qquad \alpha=x,y,z,\\
\hat{\mathcal L}_{\alpha,rs}
&=\delta_{rs}\hat{\mathfrak l}_{\alpha}
-i\hbar\Gamma_{\alpha,rs}(\boldsymbol k),\\
\Gamma_{\alpha,rs}(\boldsymbol k)
&=\boldsymbol\epsilon^{*}(\boldsymbol k,r)\cdot
\bigl[(\boldsymbol k\times\boldsymbol\nabla_{\boldsymbol k})_{\alpha}
\boldsymbol\epsilon(\boldsymbol k,s)\bigr],\qquad
\hat{\mathfrak l}_{\alpha}=-i\hbar
(\boldsymbol k\times\boldsymbol\nabla_{\boldsymbol k})_{\alpha}.
\end{align}
The normally ordered square of each component is
\begin{align}
(\hat L^{\rm obs}_{\alpha})^{2}
={}&\int d^3k\,d^3k'\sum_{r,s,t,u=1}^{2}
\hat a_{\boldsymbol k,r}^{\dagger}
\hat a_{\boldsymbol k',t}^{\dagger}
\hat{\mathcal L}^{(\boldsymbol k)}_{\alpha,rs}
\hat{\mathcal L}^{(\boldsymbol k')}_{\alpha,tu}
\hat a_{\boldsymbol k',u}\hat a_{\boldsymbol k,s}\nonumber\\
&+\int d^3k\sum_{r,s=1}^{2}
\hat a_{\boldsymbol k,r}^{\dagger}
[\hat{\mathcal L}_{\alpha}^{2}]_{rs}
\hat a_{\boldsymbol k,s}.
\end{align}
}
\textcolor{red}{Here $\hat{\mathcal L}_{\alpha}^{2}$ denotes the square of the full matrix-valued differential operator, including derivatives of the connection. The coordinate expressions below give only its ordinary differential part $\hat{\mathfrak l}_{\alpha}$.} In a Cartesian coordinate frame, these differential operators are given by
\begin{equation}
\hat{\mathfrak{l}}_{x}=-i\hbar\left(k_{y}\frac{\partial}{\partial k_{z}}-k_{z}\frac{\partial}{\partial k_{y}}\right),\ \hat{\mathfrak{l}}_{y}=-i\hbar\left(k_{z}\frac{\partial}{\partial k_{x}}-k_{x}\frac{\partial}{\partial k_{z}}\right),\ \hat{\mathfrak{l}}_{z}=\textcolor{red}{-i\hbar}\left(k_{x}\frac{\partial}{\partial k_{y}}-k_{y}\frac{\partial}{\partial k_{x}}\right).
\end{equation}

For propagating quantum pulses, it is more convenient to evaluate the photon OAM in a spherical or cylindrical coordinate frame usually. Here, we also give the expression of the differential operators in these two coordinate frames. 

Similar to the real-space counterparts, the reciprocal-space differential operators in a spherical coordinate with 
\begin{align}
k_{x} & =k\sin\theta_{k}\cos\varphi_{k},\\
k_{y} & =k\sin\theta_{k}\sin\varphi_{k},\\
k_{z} & =k\cos\theta_{k},
\end{align}
are given by
\begin{align}
\hat{\mathfrak{l}}_{x} & =
i\hbar\left(\sin\varphi_{k}\frac{\partial}{\partial\theta_{k}}+\cot\theta_{k}\cos\varphi_{k}\frac{\partial}{\partial\varphi_{k}}\right), \\
\hat{\mathfrak{l}}_{y} & = i\hbar\left(-\cos\varphi_{k}\frac{\partial}{\partial\theta_{k}}+\cot\theta_{k}\sin\varphi_{k}\frac{\partial}{\partial\varphi_{k}}\right),\\
\hat{\mathfrak{l}}_{z} & =-i\hbar\frac{\partial}{\partial\varphi_{k}}.
\end{align}
We also have the summation of their squares
\begin{equation}
\hat{\mathfrak{l}}_{x}^{2}+\hat{\mathfrak{l}}_{y}^{2}+\hat{\mathfrak{l}}_{z}^{2} =-\hbar^{2}\left[\frac{1}{\sin\theta_{k}}\frac{\partial}{\partial\theta_{k}}\sin\theta_{k}\frac{\partial}{\partial\theta_{k}}+\frac{1}{\sin^{2}\theta_{k}}\frac{\partial^{2}}{\partial\varphi_{k}^{2}}\right].
\end{equation}

In a cylindrical coordiante with
\begin{align}
k_{x} & =\rho_{k}\cos\varphi_{k},\\
k_{y} & =\rho_{k}\sin\varphi_{k},\\
k_{z} & =k_{z},
\end{align}
we have
\begin{align}
\hat{\mathfrak{l}}_{x} &=-i\hbar\left(\rho_{k}\sin\varphi_{k}\frac{\partial}{\partial k_{z}}-k_{z}\sin\varphi_{k}\frac{\partial}{\partial\rho_{k}}-\frac{k_{z}}{\rho_{k}}\cos\varphi_{k}\frac{\partial}{\partial\varphi_{k}}\right), \\
\hat{\mathfrak{l}}_{y} & = i\hbar\left(\rho_{k}\cos\varphi_{k}\frac{\partial}{\partial k_{z}}-k_{z}\cos\varphi_{k}\frac{\partial}{\partial\rho_{k}}+\frac{k_{z}}{\rho_{k}}\sin\varphi_{k}\frac{\partial}{\partial\varphi_{k}}\right),\\
\hat{\mathfrak{l}}_{z} & =-i\hbar\frac{\partial}{\partial\varphi_{k}}.
\end{align}
To evaluate the uncertainty of all components of photon OAM, we also need the the following differential operators
\begin{align}
\hat{\mathfrak{l}}_{x}^{2} = & -\hbar^{2}\left[\rho_{k}^{2}\sin^{2}\varphi_{k}\frac{\partial^{2}}{\partial k_{z}^{2}}+k_{z}^{2}\sin^{2}\varphi_{k}\frac{\partial^{2}}{\partial\rho_{k}^{2}}+\frac{k_{z}^{2}}{\rho_{k}^{2}}\cos^{2}\varphi_{k}\frac{\partial^{2}}{\partial\varphi_{k}^{2}}-\frac{k_{z}^{2}}{\rho_{k}^{2}}\sin\varphi_{k}\cos\varphi_{k}\frac{\partial}{\partial\varphi_{k}}-2\rho_{k}k_{z}\sin^{2}\varphi_{k}\frac{\partial}{\partial k_{z}}\frac{\partial}{\partial\rho_{k}}\right.\nonumber\\
 & -\rho_{k}\sin^{2}\varphi_{k}\frac{\partial}{\partial\rho_{k}}-k_{z}\sin^{2}\varphi_{k}\frac{\partial}{\partial k_{z}}+2\frac{k_{z}^{2}}{\rho_{k}}\sin\varphi_{k}\cos\varphi_{k}\frac{\partial}{\partial\rho_{k}}\frac{\partial}{\partial\varphi_{k}}-\frac{k_{z}^{2}}{\rho_{k}^{2}}\sin\varphi_{k}\cos\varphi_{k}\frac{\partial}{\partial\varphi_{k}}+\frac{k_{z}^{2}}{\rho_{k}}\cos^{2}\varphi_{k}\frac{\partial}{\partial\rho_{k}}\nonumber\\
 & \left.-2k_{z}\sin\varphi_{k}\cos\varphi_{k}\frac{\partial}{\partial k_{z}}\frac{\partial}{\partial\varphi_{k}}-\sin\varphi_{k}\cos\varphi_{k}\frac{\partial}{\partial\varphi_{k}}-k_{z}\cos^{2}\varphi_{k}\frac{\partial}{\partial k_{z}}\right],   
\end{align}
\begin{align}
\hat{\mathfrak{l}}_{y}^{2} = & -\hbar^{2}\left[\rho_{k}^{2}\cos^{2}\varphi_{k}\frac{\partial^{2}}{\partial k_{z}^{2}}+k_{z}^{2}\cos^{2}\varphi_{k}\frac{\partial^{2}}{\partial\rho_{k}^{2}} +\frac{k_{z}^{2}}{\rho_{k}^{2}}\sin^{2}\varphi_{k}\frac{\partial^{2}}{\partial\varphi_{k}^{2}}+\frac{k_{z}^{2}}{\rho_{k}^{2}}\sin\varphi_{k}\cos\varphi_{k}\frac{\partial}{\partial\varphi_{k}}-2\rho_{k}k_{z}\cos^{2}\varphi_{k}\frac{\partial}{\partial k_{z}}\frac{\partial}{\partial\rho_{k}}\right.\nonumber\\
& -\rho_{k}\cos^{2}\varphi_{k}\frac{\partial}{\partial\rho_{k}}-k_{z}\cos^{2}\varphi_{k}\frac{\partial}{\partial k_{z}}-2\frac{k_{z}^{2}}{\rho_{k}}\sin\varphi_{k}\cos\varphi_{k}\frac{\partial}{\partial\rho_{k}}\frac{\partial}{\partial\varphi_{k}}+\frac{k_{z}^{2}}{\rho_{k}^{2}}\sin\varphi_{k}\cos\varphi_{k}\frac{\partial}{\partial\varphi_{k}}+\frac{k_{z}^{2}}{\rho_{k}}\sin^{2}\varphi_{k}\frac{\partial}{\partial\rho_{k}}\nonumber\\
 & \left.+2k_{z}\sin\varphi_{k}\cos\varphi_{k}\frac{\partial}{\partial k_{z}}\frac{\partial}{\partial\varphi_{k}}+\sin\varphi_{k}\cos\varphi_{k}\frac{\partial}{\partial\varphi_{k}}-k_{z}\sin^{2}\varphi_{k}\frac{\partial}{\partial k_{z}}\right],
\end{align}
\begin{equation}
 \hat{\mathfrak{l}}_{z}^{2} = -\hbar^2\frac{\partial^2}{\partial\varphi_k^2}. 
\end{equation}
Their summation gives
\begin{equation}
\hat{\mathfrak{l}}_{x}^{2}+\hat{\mathfrak{l}}_{y}^{2}+\hat{\mathfrak{l}}_{z}^{2} =   -\hbar^{2}\left[\left(k_{z}\frac{\partial}{\partial\rho_{k}}-\rho_{k}\frac{\partial}{\partial k_{z}}\right)^{2}-k_{z}\frac{\partial}{\partial k_{z}}+\frac{k_{z}^{2}}{\rho_{k}}\frac{\partial}{\partial\rho_{k}}+\left(\frac{k_{z}^{2}}{\rho_{k}^{2}}+1\right)\frac{\partial^{2}}{\partial\varphi_{k}^{2}}\right].
\end{equation}

\vspace{0.5cm}
\noindent\textbf{Supplementary Note 3: Wave function of a quantum twisted photon pulse}
\vspace{0.5cm}

Inside a cavity, a single photon is often approximated by a single frequency mode. However, in free space, a pulse of light can simultaneously contain one photon and many frequencies. \textcolor{red}{An attenuated laser pulse produces a coherent state (which is not a single-photon Fock state even when its mean photon number is one), whereas a single-photon Fock-state pulse can be produced by emission from a single atom or quantum dot.}
\textcolor{red}{An ideal photon-number-resolving measurement of a single-photon Fock-state pulse yields one photon in the absence of loss.} The various frequencies can be separated using a tunable spectral filter and the probability distribution can be constructed through repeated measurements. We note that, \textcolor{red}{unlike an idealized plane wave or stationary cavity mode, a normalizable free-space single-photon pulse of finite duration has nonzero spectral bandwidth.} Thus, the Fock-state single-photon pulse must be a superposition of Fock states $|1_k\rangle$ of different modes $\sum_k \xi_k |1_k\rangle$. Here, we give the formalism of quantum states of light pulses in the continuum-mode pulse limit.

The single-photon wave-packet creation operator for an arbitrary quantum twisted photon pulse  is defined as
\begin{equation}
\hat{a}_{\xi\lambda}^{\dagger}=\int d^{3}k\xi_{\lambda}(\boldsymbol{k})\hat{a}_{\boldsymbol{k}\lambda}^{\dagger},
\end{equation}
which is a coherent superposition of plane waves with fixed \textcolor{red}{helicity} $\lambda$. In this work, we only focus on twisted photon pulse, thus the spectral amplitude function (SAF) will have a helical wavefront phase. Generally, the SAF of a twisted pulse with \textcolor{red}{a definite azimuthal mode index} can be expressed as
\begin{equation}
\xi_{\lambda}(\boldsymbol{k})=\frac{1}{\sqrt{2\pi}}\eta_{\lambda}(k_{z},\rho_{k})e^{im\varphi_{k}},
\end{equation}
where $\eta_{\lambda}(k_{z},\rho_{k})$ is determined by the the pulse shape and pulse type as shown in the following. The phase $\exp (i m\varphi_k)$ with an integer $m$ \textcolor{red}{labels the azimuthal winding; the physical OAM also includes the frame connection (Supplementary Note~2).} To guarantee that $\hat{a}_{\xi\lambda}^{\dagger}$ obeys the bosonic commutation relation $[\hat{a}_{\xi\lambda},\hat{a}_{\xi\lambda}^{\dagger}]=1$, the SAF should satisfy the normalization condition
\begin{equation}
  \int d^{3}k\left|\xi_{\lambda}(\boldsymbol{k})\right|^{2}=1.  
\end{equation}

\subsection*{Fock-state and coherent-state photon pulses}
The wave-packet creation operator $\hat{a}_{\xi\lambda}^{\dagger}$ can be treated as a normal ladder operator of a harmonic oscillator. Thus, we can construct arbitrary photon-pulse state~\cite{loudon2000quantum}. We first construct the the $n$-photon Fock-state pulse
\begin{equation}
\left|n_{\xi\lambda}\right\rangle =\frac{1}{\sqrt{n!}}\left(\hat{a}_{\xi\lambda}^{\dagger}\right)^{n}\left|0\right\rangle.
\end{equation}
Using the commutation relation $[\hat{a}_{\xi\lambda},\hat{a}_{\xi\lambda}^{\dagger}]=1$, we can verify that this quantum state satisfies the orthonormal relation
\begin{equation}
\langle n_{\xi\lambda}|n'_{\xi\lambda}\rangle=\delta_{nn'}.   
\end{equation}
Using the relations
\begin{align}
[\hat{a}_{\boldsymbol{k}',\lambda'},\left(\hat{a}_{\xi\lambda}^{\dagger}\right)^{n}] & =n\xi_{\lambda'}(\boldsymbol{k}')\delta_{\lambda\lambda'}\left(\hat{a}_{\xi\lambda}^{\dagger}\right)^{n-1},\label{eq:CR1}\\{}
[\hat{a}_{\boldsymbol{k}',\lambda'}^{\dagger},\left(\hat{a}_{\xi\lambda}\right)^{n}] & =-n\xi_{\lambda'}^{*}(\boldsymbol{k}')\delta_{\lambda\lambda'}\left(\hat{a}_{\xi\lambda}\right)^{n-1},\label{eq:CR2}
\end{align}
we have
\begin{align}
\hat{a}_{\boldsymbol{k}',\lambda'}\left|n_{\xi \lambda}\right\rangle  & =\sqrt{n}\xi_{\lambda'}(\boldsymbol{k}')\left|(n-1)_{\xi \lambda}\right\rangle \delta_{\lambda'\lambda},\\
\left\langle n_{\xi \lambda}\right|\hat{a}_{\boldsymbol{k}',\lambda'}^{\dagger} & =\sqrt{n}\xi_{\lambda'}^{*}(\boldsymbol{k}')\left\langle (n-1)_{\xi \lambda}\right|\delta_{\lambda'\lambda}.
\end{align}
We can also verify that the photon number in this $n$-photon Fock-state pulse is exact $n$, i.e.,
\begin{equation}
\langle n_{\xi\lambda}|\int d^3k\sum_{\lambda'=\pm}\hat{a}_{\boldsymbol{k},\lambda'}^{\dagger}\hat{a}_{\boldsymbol{k},\lambda'}|n_{\xi\lambda}\rangle = n.   
\end{equation}

Similarly, we can construct a coherent-state pulse
\begin{equation}
\left|\alpha_{\xi\lambda}\right\rangle =\exp\left(\alpha\hat{a}_{\xi\lambda}^{\dagger}-\frac{1}{2}\bar{n}\right)\left|0\right\rangle =e^{-\bar{n}/2}\sum_{n=0}^{\infty}\frac{\alpha^{n}}{n!}\left(\hat{a}_{\xi\lambda}^{\dagger}\right)^{n}\left|0\right\rangle=e^{-\bar{n}/2}\sum_{n=0}^{\infty}\frac{\alpha^{n}}{\sqrt{n!}}\left|n_{\xi\lambda}\right\rangle ,
\end{equation}
where $\bar{n}=|\alpha|^{2}$ is the mean photon number in the coherent-state pulse. We can check that this state is well normalized 
\begin{equation}
\left\langle \alpha_{\xi\lambda}\mid\alpha_{\xi\lambda}\right\rangle =e^{-\bar{n}}\sum_{m,n=0}^{\infty}\frac{|\alpha|^{2n}}{\sqrt{m!n!}}\left\langle m_{\xi\lambda}\mid n_{\xi\lambda}\right\rangle =\textcolor{red}{e^{-\bar{n}}\sum_{n=0}^{\infty}\frac{\bar{n}^{n}}{n!}}=1.
\end{equation}
It also has the coherent-state property
\begin{equation}
\hat{a}_{\boldsymbol{k}',\lambda'}\left|\alpha_{\xi\lambda}\right\rangle =\alpha\xi_{\lambda}(\boldsymbol{k}')\delta_{\lambda'\lambda}\left|\alpha_{\xi\lambda}\right\rangle,
\end{equation}

\subsection*{Spectral amplitude function}
In this subsection, we give the SAF for the Bessel pulse and Laguerre-Gauss pulse, which are the most common twisted photon pulses used in experiments. In the main text, we only studied the Bessel pulse.

We first look at the Bessel pulse. The SAF of a Bessel pulse with a Gaussian envelop can be express $\eta_{\lambda}(k_{z},\rho_{k})$
as the product of two Gaussian functions
\begin{align}
\eta_{\lambda}(k_{z},\rho_{k})=&\left(\frac{2\sigma_{z}^{2}}{\pi}\right)^{1/4}\exp\left[-\sigma_{z}^{2}(k_{z}-k_{z,c})^{2}\right] \nonumber \\
&\times\left(\frac{2\sigma_{\rho}^{2}}{\pi k_{\perp,c}^{2}}\right)^{1/4}\exp\left[-\sigma_{\rho}^{2}(\rho_{k}-k_{\perp,c})^{2}\right].
\end{align}
The first Gaussian function with width $1/\sigma_z$ and center wave vector $k_{z,c}$ characterize the envelope of the pulse in the propagating axis. The pulse length on $z$-axis in real space is given by $\sigma_{z}=c\tau_{p}$ ($\tau_{p}$ the
pulse length in time). The second Gaussian function with width
$1/\sigma_{\rho}$ and center value $k_{\perp,c}=k_{z,c}\tan\theta_{c}$ reflects the fact that a single-frequency Bessel beam is the superposition of all plane waves on a cone in $k$-space. \textcolor{red}{Here, for simplicity, we choose the two Gaussian functions to have the same ratio between the central wave number and the spectral width}, i.e., $k_{z,c}\sigma_z= k_{\perp,c}\sigma_{\rho}\equiv C_0$.  In the narrow bandwidth limit $C_0\gg 1$, our defined SAF is well normalized
\begin{align}
 \int d^3k\textcolor{red}{|\xi_{\lambda}(\boldsymbol{k})|^{2}}=& \int_{-\infty}^{\infty}dk_{z}\int_{0}^{\infty}\rho_{k}d\rho_{k}\int_{0}^{2\pi}d\varphi_{k}\frac{1}{2\pi}\eta_{\lambda}^{2}(k_{z},\rho_{k})\nonumber \\
= & \left(\frac{2\sigma_{z}^{2}}{\pi}\right)^{1/2}\int_{-\infty}^{\infty}dk_{z}e^{-2\sigma_{z}^{2}(k_{z}-k_{z,c})^{2}}\times\left(\frac{2\sigma_{\rho}^{2}}{\pi}\right)^{1/2}\int_{0}^{\infty}\frac{\rho_{k}}{k_{\perp,c}}e^{-2\sigma_{\rho}^{2}(\rho_{k}-k_{\perp,c})^{2}}d\rho_{k}\\
\approx & \left(\frac{2\sigma_{z}^{2}}{\pi}\right)^{1/2}\int_{-\infty}^{\infty}dk_{z}e^{-2\sigma_{z}^{2}(k_{z}-k_{z,c})^{2}}\times\left(\frac{2\sigma_{\rho}^{2}}{\pi}\right)^{1/2}\int_{-\infty}^{\infty}e^{-2\sigma_{\rho}^{2}(\rho_{k}-k_{\perp,c})^{2}}d\rho_{k}\\
= & 1.
\end{align}

The SAF of a Laguerre-Gaussian (LG) pulse with a Gaussian evelope is given be
\begin{align}
\xi_{pm\lambda}(\boldsymbol{k})= &\mathcal{N}^{-1/2}\textcolor{red}{\left(\frac{4\sigma_{z}^{2}}{\pi}\right)^{1/4}}e^{-2\sigma_{z}^{2}(k_{z}-k_{z,c})^{2}} \times\sqrt{\frac{1}{2\pi}}\rho_{k}^{2p+|m|}e^{im\varphi_{k}}\exp\left[-\frac{w_{0}^{2}}{4}(1+i\zeta)\rho_{k}^{2}\right],
\end{align}
Similarly, the first Gaussian function with width $1/\sigma_z$ and center wave vector $k_{z,c}$ characterize the envelope of the pulse in the propagating axis. The second complicated function with integer numbers $m$ and $p$ characterize the amplitude function of the elegant LG mode in $k$-space~\cite{enderlein2004unified}. Here, $w_0$ is the pulse waist in $xy$-plane and $\zeta = z/z_R$ is the reduced coordinate with respect to the Rayleigh lenght $z_R=k_z w_0^2/2$. The normalization factor $\mathcal{N}$ is given by
\begin{equation}
\mathcal{N}=\frac{\Gamma(2p+|m|+1)}{2(w_{0}^{2}/2)^{2p+|m|+1}}.
\end{equation}

\vspace{0.5cm}
\noindent\textbf{Supplementary Note 4: Quantum statistics of the angular momenta of twisted photon pulses}
\vspace{0.5cm}

In this section, we give the details of the evaluation of the mean value and quantum uncertainty of the angular momentum of Fock-state and coherent-state twisted photon pulses. 

\subsection*{Spin angular momentum of twisted photon pulses}
We first study the photon spin for a Fock-state Bessel pulse. It mean value is given by
\begin{align}
\left\langle n_{\xi\lambda}\right|\hat{\boldsymbol{S}}^{{\rm obs}}\left|n_{\xi\lambda}\right\rangle & =  \hbar\int d^{3}k\left\langle n_{\xi\lambda}\right|(\hat{a}_{\boldsymbol{k},+}^{\dagger}\hat{a}_{\boldsymbol{k},+}-\hat{a}_{\boldsymbol{k},-}^{\dagger}\hat{a}_{\boldsymbol{k},-})\left|n_{\xi\lambda}\right\rangle \boldsymbol{e}(\boldsymbol{k},3)\\
& = \hbar n\lambda\int d^{3}k\left|\xi_{\lambda}(\boldsymbol{k})\right|^{2} \boldsymbol{e}(\boldsymbol{k},3)\\
& \approx \frac{\hbar n\lambda}{2\pi}\int_{-\infty}^{\infty} dk_{z}\int_0^{\infty}\rho_{k}d\rho_{k}\int_0^{2\pi} d\varphi_{k}\mid\eta_{\lambda}(k_{z},\rho_{k})\mid^{2}\left(\sin\theta_{c}\cos\varphi_{k},\sin\theta_{c}\sin\varphi_{k},\cos\theta_{c}\right)\\
 & =\hbar n\lambda\left(0,0,\cos\theta_{c}\right),
\end{align}
where in the thirst step, we have used the approximation the unit vector
$\boldsymbol{e}(\boldsymbol{k},3)=\boldsymbol{k}/|\boldsymbol{k}|\approx\left(\sin\theta_{c}\cos\varphi_{k},\sin\theta_{c}\sin\varphi_{k},\cos\theta_{c}\right)$. Now, we calculate the mean values of $(\hat{S}^{\rm obs}_{x})^{2}$, $(\hat{S}^{\rm obs}_{y})^{2}$, and
$(\hat{S}^{\rm obs}_{z})^{2}$ separately,
\begin{align}
\left\langle n_{\xi\lambda}\right|(\hat{S}^{\rm obs}_{z})^{2}\left|n_{\xi\lambda}\right\rangle  & \approx\left\langle n_{\xi\lambda}\right|\left[\hbar\int d^{3}k(\hat{a}_{\boldsymbol{k},+}^{\dagger}\hat{a}_{\boldsymbol{k},+}-\hat{a}_{\boldsymbol{k},-}^{\dagger}\hat{a}_{\boldsymbol{k},-})\cos\theta_{c}\right]^{2}\left|n_{\xi\lambda}\right\rangle \\
 & =\left\langle n_{\xi\lambda}\right|\hbar^{2}\cos^{2}\theta_{c}\int d^{3}k\int d^{3}k'\left[\hat{a}_{\boldsymbol{k},\lambda}^{\dagger}\hat{a}_{\boldsymbol{k}',\lambda}^{\dagger}\hat{a}_{\boldsymbol{k}',\lambda}\hat{a}_{\boldsymbol{k},\lambda}+\delta^{3}(\boldsymbol{k}-\boldsymbol{k}')\hat{a}_{\boldsymbol{k},\lambda}^{\dagger}\hat{a}_{\boldsymbol{k},\lambda}\right]\left|n_{\xi\lambda}\right\rangle \\
 & = \hbar^{2}\cos^{2}\theta_{c}\left[n(n-1)\int d^{3}k\left|\xi_{\lambda}(\boldsymbol{k})\right|^{2}\int d^{3}k'\left|\xi_{\lambda}(\boldsymbol{k'})\right|^{2}+n\int d^{3}k\left|\xi_{\lambda}(\boldsymbol{k})\right|^{2}\right]= n^{2}\hbar^{2}\cos^{2}\theta_{c}.
\end{align}
\begin{align}
\left\langle n_{\xi\lambda}\right|(\hat{S}^{\rm obs}_{x})^{2}\left|n_{\xi\lambda}\right\rangle
\approx & \left\langle n_{\xi\lambda}\right|\left[\hbar\int d^{3}k(\hat{a}_{\boldsymbol{k},+}^{\dagger}\hat{a}_{\boldsymbol{k},+}-\hat{a}_{\boldsymbol{k},-}^{\dagger}\hat{a}_{\boldsymbol{k},-})\sin\theta_{c}\cos\varphi_{k}\right]^{2}\left|n_{\xi\lambda}\right\rangle \\
= & \left\langle n_{\xi\lambda}\right|\hbar^{2}\sin^{2}\theta_{c}\int\cos\varphi_{k}d^{3}k\int\cos\varphi_{k}^{\prime}d^{3}k'\left[\hat{a}_{\boldsymbol{k},\lambda}^{\dagger}\hat{a}_{\boldsymbol{k}',\lambda}^{\dagger}\hat{a}_{\boldsymbol{k}',\lambda}\hat{a}_{\boldsymbol{k},\lambda}+\delta^{3}(\boldsymbol{k}-\boldsymbol{k}')\hat{a}_{\boldsymbol{k},\lambda}^{\dagger}\hat{a}_{\boldsymbol{k},\lambda}\right]\left|n_{\xi\lambda}\right\rangle \\
= & \frac{1}{2}n\hbar^{2}\sin^{2}\theta_{c}.
\end{align}
\begin{align}
\left\langle n_{\xi\lambda}\right|(\hat{S}^{\rm obs}_{y})^{2}\left|n_{\xi\lambda}\right\rangle
\approx & \left\langle n_{\xi\lambda}\right|\left[\hbar\int d^{3}k(\hat{a}_{\boldsymbol{k},+}^{\dagger}\hat{a}_{\boldsymbol{k},+}-\hat{a}_{\boldsymbol{k},-}^{\dagger}\hat{a}_{\boldsymbol{k},-})\sin\theta_{c}\sin\varphi_{k}\right]^{2}\left|n_{\xi\lambda}\right\rangle \\
= & \left\langle n_{\xi\lambda}\right|\hbar^{2}\sin^{2}\theta_{c}\int\sin\varphi_{k}d^{3}k\int\sin\varphi_{k}^{\prime}d^{3}k'\left[\hat{a}_{\boldsymbol{k},\lambda}^{\dagger}\hat{a}_{\boldsymbol{k}',\lambda}^{\dagger}\hat{a}_{\boldsymbol{k}',\lambda}\hat{a}_{\boldsymbol{k},\lambda}+\delta^{3}(\boldsymbol{k}-\boldsymbol{k}')\hat{a}_{\boldsymbol{k},\lambda}^{\dagger}\hat{a}_{\boldsymbol{k},\lambda}\right]\left|n_{\xi\lambda}\right\rangle \\
= & \frac{1}{2}n\hbar^{2}\sin^{2}\theta_{c}.
\end{align}
Thus, the standard derivations of the spin of an $n$-photon Fock-state Bessel pulse are given by
\begin{align}
\Delta\hat{S}^{\rm obs}_{x}  =\Delta\hat{S}^{\rm obs}_{y}=\hbar\sqrt{n/2}\left|\sin\theta_{c}\right|,\ 
\Delta\hat{S}^{\rm obs}_{z} \textcolor{red}{\approx}0.
\end{align}

Similarly, we can calculate the mean value and the quantum uncertainty of the spin of a coherent-state Bessel pulse:
\begin{equation}
\left\langle \alpha_{\xi\lambda}\right|\hat{\boldsymbol{S}}^{\rm obs}\left|\alpha_{\xi\lambda}\right\rangle =  \hbar\int d^{3}k\left\langle \alpha_{\xi\lambda}\right|(\hat{a}_{\boldsymbol{k},+}^{\dagger}\hat{a}_{\boldsymbol{k},+}-\hat{a}_{\boldsymbol{k},-}^{\dagger}\hat{a}_{\boldsymbol{k},-})\left|\alpha_{\xi\lambda}\right\rangle \boldsymbol{e}(\boldsymbol{k},3)
= \hbar\bar{n}\lambda\left(0,0,\cos\theta_{c}\right),
\end{equation}
where $\bar{n}=|\alpha|^{2}$ is the mean photon number of the pulse. The mean values of $\hat{S}_{M,x}^{2}$, $\hat{S}_{M,y}^{2}$, and
$\hat{S}_{M,z}^{2}$ are given by
\begin{align}
\left\langle \alpha_{\xi\lambda}\right|(\hat{S}^{\rm obs}_{z})^{2}\left|\alpha_{\xi\lambda}\right\rangle  & \approx\left\langle \alpha_{\xi\lambda}\right|\left[\hbar\int d^{3}k(\hat{a}_{\boldsymbol{k},+}^{\dagger}\hat{a}_{\boldsymbol{k},+}-\hat{a}_{\boldsymbol{k},-}^{\dagger}\hat{a}_{\boldsymbol{k},-})\cos\theta_{c}\right]^{2}\left|\alpha_{\xi\lambda}\right\rangle \\
 & =\left\langle \alpha_{\xi\lambda}\right|\hbar^{2}\cos^{2}\theta_{c}\int d^{3}k\int d^{3}k'\left[\hat{a}_{\boldsymbol{k},\lambda}^{\dagger}\hat{a}_{\boldsymbol{k}',\lambda}^{\dagger}\hat{a}_{\boldsymbol{k}',\lambda}\hat{a}_{\boldsymbol{k},\lambda}+\delta^{3}(\boldsymbol{k}-\boldsymbol{k}')\hat{a}_{\boldsymbol{k},\lambda}^{\dagger}\hat{a}_{\boldsymbol{k},\lambda}\right]\left|\alpha_{\xi\lambda}\right\rangle \\
 & =\bar{n}(\bar{n}+1)\hbar^{2}\cos^{2}\theta_{c}.
\end{align}
\begin{align}
 \left\langle \alpha_{\xi\lambda}\right|(\hat{S}^{\rm obs}_{x})^{2}\left|\alpha_{\xi\lambda}\right\rangle
\approx & \left\langle \alpha_{\xi\lambda}\right|\left[\hbar\int d^{3}k(\hat{a}_{\boldsymbol{k},+}^{\dagger}\hat{a}_{\boldsymbol{k},+}-\hat{a}_{\boldsymbol{k},-}^{\dagger}\hat{a}_{\boldsymbol{k},-})\sin\theta_{c}\cos\varphi_{k}\right]^{2}\left|\alpha_{\xi\lambda}\right\rangle \\
= & \left\langle \alpha_{\xi\lambda}\right|\hbar^{2}\sin^{2}\theta_{c}\int\cos\varphi_{k}d^{3}k\int\cos\varphi_{k}^{\prime}d^{3}k'\left[\hat{a}_{\boldsymbol{k},\lambda}^{\dagger}\hat{a}_{\boldsymbol{k}',\lambda}^{\dagger}\hat{a}_{\boldsymbol{k}',\lambda}\hat{a}_{\boldsymbol{k},\lambda}+\delta^{3}(\boldsymbol{k}-\boldsymbol{k}')\hat{a}_{\boldsymbol{k},\lambda}^{\dagger}\hat{a}_{\boldsymbol{k},\lambda}\right]\left|\alpha_{\xi\lambda}\right\rangle \\
= & \frac{1}{2}\bar{n}\hbar^{2}\sin^{2}\theta_{c}.
\end{align}
\begin{align}
\left\langle \alpha_{\xi\lambda}\right|(\hat{S}^{\rm obs}_{y})^{2}\left|\alpha_{\xi\lambda}\right\rangle 
\approx & \left\langle \alpha_{\xi\lambda}\right|\left[\hbar\int d^{3}k(\hat{a}_{\boldsymbol{k},+}^{\dagger}\hat{a}_{\boldsymbol{k},+}-\hat{a}_{\boldsymbol{k},-}^{\dagger}\hat{a}_{\boldsymbol{k},-})\sin\theta_{c}\sin\varphi_{k}\right]^{2}\left|\alpha_{\xi\lambda}\right\rangle \\
= & \left\langle \alpha_{\xi\lambda}\right|\hbar^{2}\sin^{2}\theta_{c}\int\sin\varphi_{k}d^{3}k\int\sin\varphi_{k}^{\prime}d^{3}k'\left[\hat{a}_{\boldsymbol{k},\lambda}^{\dagger}\hat{a}_{\boldsymbol{k}',\lambda}^{\dagger}\hat{a}_{\boldsymbol{k}',\lambda}\hat{a}_{\boldsymbol{k},\lambda}+\delta^{3}(\boldsymbol{k}-\boldsymbol{k}')\hat{a}_{\boldsymbol{k},\lambda}^{\dagger}\hat{a}_{\boldsymbol{k},\lambda}\right]\left|\alpha_{\xi\lambda}\right\rangle \\
= & \textcolor{red}{\frac{1}{2}\bar{n}\hbar^{2}\sin^{2}\theta_{c}}.
\end{align}
Thus, the standard derivations of the spin of \textcolor{red}{a} coherent-state Bessel pulse are given by
\begin{equation}
\Delta\hat{S}^{\rm obs}_{x} =\Delta\hat{S}^{\rm obs}_{y}=\hbar\sqrt{\bar{n}/2}\left|\sin\theta_{c}\right|,\ 
\Delta\hat{S}^{\rm obs}_{z} =\textcolor{red}{\hbar\sqrt{\bar n}}\left|\cos\theta_{c}\right|.
\end{equation}

\subsection*{Orbital angular momentum of twisted photon pulses}
{\color{red}
In this subsection we evaluate the observable OAM, including the polarization-frame connection derived in Supplementary Notes 1 and 2. We use the local spherical helicity frame in which
\begin{align}
\boldsymbol{\mathcal A}_{\lambda}(\boldsymbol k)
&=\frac{\lambda\cos\theta_k}{k\sin\theta_k}\boldsymbol e_{\varphi_k},\nonumber\\
\hat{\mathcal L}^{(\lambda)}_i
&=-i\hbar\bigl[\boldsymbol k\times
 (\boldsymbol\nabla_{\boldsymbol k}-i\boldsymbol{\mathcal A}_{\lambda})\bigr]_i,
\end{align}
and retain the state convention of Supplementary Note 3,
$\xi_{\lambda}(\boldsymbol k)
=\eta_{\lambda}(k_z,\rho_k)e^{im\varphi_k}/\sqrt{2\pi}$,
with real, normalized $\eta_{\lambda}$. In this frame,
$\hat{\mathcal L}^{(\lambda)}_z=-i\hbar\partial_{\varphi_k}
-\hbar\lambda\cos\theta_k$, where
$\cos\theta_k=k_z/\sqrt{k_z^2+\rho_k^2}$.
}
The mean value of $\hat{L}^{\rm obs}_z$ for a Fock-state twisted pulse is given by
{\color{red}
\begin{align}
\left\langle n_{\xi\lambda}\right|\hat L_z^{\rm obs}
\left|n_{\xi\lambda}\right\rangle
&=n\hbar\int d^3k\,|\xi_{\lambda}(\boldsymbol k)|^2
 (m-\lambda\cos\theta_k)\nonumber\\
&=n\hbar(m-\lambda\mu)
\approx n\hbar(m-\lambda\cos\theta_c),
\qquad \mu\equiv\int d^3k\,|\xi_{\lambda}(\boldsymbol k)|^2
\cos\theta_k.\label{eq:Lz1}
\end{align}
The connection contribution makes the observable OAM dependent on
both the azimuthal winding $m$ and the helicity $\lambda$; the sum with
$\langle S_z^{\rm obs}\rangle=n\hbar\lambda\mu$ gives
$\langle J_z^{\rm obs}\rangle=nm\hbar$ in the present helicity frame.
For an azimuthally symmetric cone spectrum,
\begin{equation}
\langle\hat L_x^{\rm obs}\rangle
=\langle\hat L_y^{\rm obs}\rangle=0.
\end{equation}
The corresponding longitudinal second moment and variance are
\begin{align}
\langle(\hat L_z^{\rm obs})^2\rangle
&=n(n-1)\hbar^2(m-\lambda\mu)^2
+n\hbar^2\int d^3k\,|\xi_\lambda|^2
 (m-\lambda\cos\theta_k)^2,\\
(\Delta\hat L_z^{\rm obs})^2
&=n\hbar^2\left(\langle\cos^2\theta_k\rangle-\mu^2\right)
\approx 0,
\end{align}
where $\langle f\rangle\equiv\int d^3k\,|\xi_\lambda|^2 f$;
the last approximation assumes a narrow cone.

The two transverse OAM variances must be calculated with the full
connection-aware operators. Writing
$K\equiv k_z\partial_{\rho_k}-\rho_k\partial_{k_z}$,
we obtain directly from $\|\hat{\mathcal L}^{(\lambda)}_{x,y}\xi_\lambda\|^2$
\begin{align}
\left\langle(\hat L_x^{\rm obs})^2\right\rangle
=\left\langle(\hat L_y^{\rm obs})^2\right\rangle
=\frac{n\hbar^2}{2}
\int_{-\infty}^{\infty}\!dk_z\int_0^\infty\!\rho_kd\rho_k\,
\left[(K\eta_\lambda)^2+
\frac{k_z^2}{\rho_k^2}
(m-\lambda\cos\theta_k)^2\eta_\lambda^2\right].
\label{eq:note4-Lxy2}
\end{align}
In particular, retaining the polarization connection amounts to
replacing the bare azimuthal contribution
$m^2 k_z^2/\rho_k^2$ by
$(m-\lambda\cos\theta_k)^2k_z^2/\rho_k^2$.
The Gaussian Bessel envelope introduced in Supplementary Note 3 is
understood as a narrow-cone approximation with a smooth cutoff away
from the frame singularities (in particular $\rho_k=0$): an uncut
Gaussian tail generally makes the exact $1/\rho_k^2$ moment divergent.
For $C_0\gg1$, with the cutoff negligible in the main spectral cone,
Eq.~\eqref{eq:note4-Lxy2} gives, to the same narrow-band accuracy as
Supplementary Note 3,
\begin{align}
\left\langle(\hat L_x^{\rm obs})^2\right\rangle
=\left\langle(\hat L_y^{\rm obs})^2\right\rangle
\approx\frac{n\hbar^2}{2}\Bigg[
&C_0^2(\tan^2\theta_c+\cot^2\theta_c)
+\frac14\tan^2\theta_c+\frac34\cot^2\theta_c-1\nonumber\\
&+(m-\lambda\cos\theta_c)^2\cot^2\theta_c
\Bigg].
\end{align}
The observable transverse photon OAM has the noncanonical algebra
$[\hat L^{\rm obs}_x,\hat L^{\rm obs}_y]
=i\hbar(\hat L^{\rm obs}_z-\hat S^{\rm obs}_z)$,
so the appropriate Robertson bound is
\begin{equation}
\Delta\hat L_x^{\rm obs}\,\Delta\hat L_y^{\rm obs}
\geq\frac{\hbar}{2}
\bigl|\langle\hat L_z^{\rm obs}-\hat S_z^{\rm obs}\rangle\bigr|
=\frac{n\hbar^2}{2}|m-2\lambda\mu|.
\end{equation}
The other two bounds have vanishing right-hand sides for an
azimuthally symmetric cone spectrum. For the corresponding coherent
pulse, the transverse variances follow from the same one-photon
second moments with $n$ replaced by $\bar n$.
}

\vspace{0.5cm}
\noindent\textbf{Supplementary Note 5: Quantum spin texture of Bessel photon pulse}
\vspace{0.5cm}

In this section, we give the details about the quantum properties of the photon spin texture, i.e., the photon spin density,
\begin{align}
\hat{\boldsymbol{s}}^{\rm obs}(\boldsymbol{r},t) & =\varepsilon_{0}\hat{\boldsymbol{E}}_{\perp}(\boldsymbol{r},t)\times\hat{\boldsymbol{A}}_{\perp}(\boldsymbol{r},t)\\
 & \simeq \frac{-i\hbar}{2(2\pi)^{3}}\sum_{\lambda'\lambda''}\int d^{3}k'\int d^{3}k''\left(\sqrt{\frac{\omega'}{\omega''}}+\sqrt{\frac{\omega''}{\omega'}}\right)\hat{a}_{\boldsymbol{k}',\lambda'}^{\dagger}\hat{a}_{\boldsymbol{k}'',\lambda''}e^{-i[(\boldsymbol{k}'-\boldsymbol{k}'')\cdot\boldsymbol{r}-(\omega'-\omega'')t]}\boldsymbol{e}^{*}(\boldsymbol{k}',\lambda')\times\boldsymbol{e}(\boldsymbol{k}'',\lambda'').
\end{align}
In the narrow bandwidth limit, we have $\omega'\approx \omega''$. The number-nonconserving contributions are neglected at this order. The spin densities of an $n$-photon Fock-state and a coherent-state pulses are given by
\begin{align}
\left\langle n_{\xi\lambda}\right|\hat{\boldsymbol{s}}^{\rm obs}(\boldsymbol{r},t)\left|n_{\xi\lambda}\right\rangle \approx & \frac{-in\hbar}{(2\pi)^{3}}\int d^{3}k'\int d^{3}k''\xi_{\lambda}^{*}(\boldsymbol{k}')\xi_{\lambda}(\boldsymbol{k}'')e^{-i[(\boldsymbol{k}'-\boldsymbol{k}'')\cdot\boldsymbol{r}-(\omega'-\omega'')t]}\boldsymbol{e}^{*}(\boldsymbol{k}',\lambda)\times\boldsymbol{e}(\boldsymbol{k}'',\lambda),
\end{align}
and
\begin{align}
\left\langle \alpha_{\xi\lambda}\right|\hat{\boldsymbol{s}}^{\rm obs}(\boldsymbol{r},t)\left|\alpha_{\xi\lambda}\right\rangle \approx & \frac{-i\bar{n}\hbar}{(2\pi)^{3}}\int d^{3}k'\int d^{3}k''\xi_{\lambda}^{*}(\boldsymbol{k}')\xi_{\lambda}(\boldsymbol{k}'')e^{-i[(\boldsymbol{k}'-\boldsymbol{k}'')\cdot\boldsymbol{r}-(\omega'-\omega'')t]}\boldsymbol{e}^{*}(\boldsymbol{k}',\lambda)\times\boldsymbol{e}(\boldsymbol{k}'',\lambda),
\end{align}
respectively. 

\begin{figure}
\includegraphics[width=8cm]{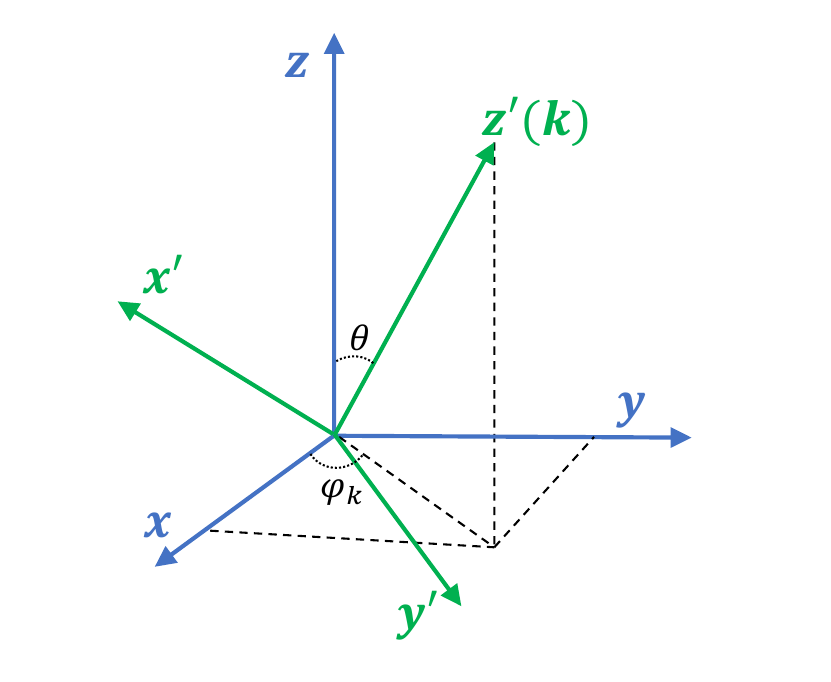}
\caption{\label{fig:S1} Relation between the wave-vector-dependent coordinate frame  (the green arrows)  and the fixed lab coordinate (the blue arrows).}
\end{figure}

Using $\boldsymbol{r}=(\rho\cos\varphi,\rho\sin\varphi,z)$
and $\boldsymbol{k}=(\rho_{k}\cos\varphi_{k},\rho_{k}\sin\varphi_{k},k_{z})$,
we can expand the phase factor as
\begin{equation}
e^{-i[(\boldsymbol{k}'-\boldsymbol{k}'')\cdot\boldsymbol{r}-(\omega'-\omega'')t]}=\exp\left\{ i\left[(\omega'-\omega^{\prime\prime})t-(k_{z}^{\prime}-k_{z}^{\prime\prime})z-\rho\rho_{k}^{\prime}\cos(\varphi-\varphi_{k}^{\prime})+\rho\rho_{k}^{\prime\prime}\cos(\varphi-\varphi_{k}^{\prime\prime})\right]\right\} .
\end{equation}
We can also expand $\boldsymbol{k}$-dependent polarization unit vectors $\boldsymbol{e}(\boldsymbol{k},\lambda)$ in the fixed lab coordinate
frame (see Fig.~\ref{fig:S1}),
\begin{equation}
\boldsymbol{e}(\boldsymbol{k},\lambda)= e^{-i\lambda\varphi_{k}}\cos^{2}\frac{\theta_{c}}{2}\boldsymbol{e}_{\lambda}-e^{i\lambda\varphi_{k}}\sin^{2}\frac{\theta_{c}}{2}\boldsymbol{e}_{-\lambda}-\frac{1}{\sqrt{2}}\sin\theta_{c}\boldsymbol{e}_{z},
\end{equation}
where $\boldsymbol{e}_{\lambda}=\left(\boldsymbol{e}_x+i\lambda \boldsymbol{e}_y\right)/\sqrt{2}$.
Thus 
\begin{align}
\boldsymbol{e}^{*}(\boldsymbol{k}',\lambda)\times\boldsymbol{e}(\boldsymbol{k}'',\lambda)= & \frac{i\lambda}{\sqrt{2}}\left[e^{-i\lambda\varphi_{k}^{\prime\prime}}\sin\theta_{c}\cos^{2}\frac{\theta_{c}}{2}+e^{-i\lambda\varphi_{k}^{\prime}}\sin\theta_{c}\sin^{2}\frac{\theta_{c}}{2}\right]\boldsymbol{e}_{\lambda}\\
 & +\frac{i\lambda}{\sqrt{2}}\left[e^{i\lambda\varphi_{k}^{\prime\prime}}\sin\theta_{c}\sin^{2}\frac{\theta_{c}}{2}+e^{i\lambda\varphi_{k}^{\prime}}\cos^{2}\frac{\theta_{c}}{2}\sin\theta_{c}\right]\boldsymbol{e}_{-\lambda}\\
 & +i\lambda\left[e^{i\lambda(\varphi_{k}^{\prime}-\varphi_{k}^{\prime\prime})}\cos^{4}\frac{\theta_{c}}{2}-e^{-i\lambda(\varphi_{k}^{\prime}-\varphi_{k}^{\prime\prime})}\sin^{4}\frac{\theta_{c}}{2}\right]\boldsymbol{e}_{z},
\end{align}
where we have used relations $\boldsymbol{e}_{\lambda}\times\boldsymbol{e}_{\lambda}^{*}=-i\lambda\boldsymbol{e}_{z}$,
$\boldsymbol{e}_{z}\times\boldsymbol{e}_{\lambda}^{*}=i\lambda\boldsymbol{e}_{-\lambda}$,
and $\boldsymbol{e}_{z}\times\boldsymbol{e}_{\lambda}=-i\lambda\boldsymbol{e}_{\lambda}$.

For convenience, we define the following amplitude function
\begin{equation}
   \psi_{\lambda}(\boldsymbol{r},t)= \frac{1}{\sqrt{(2\pi)^{3}}}\int d^{3}k\xi_{\lambda}(\boldsymbol{k})e^{i[\boldsymbol{k}\cdot\boldsymbol{r}-\omega t}+\lambda\varphi_{k}],
\end{equation}
which is the effective wave function of the photon pulse in real space.
For a Bessel pulse, we have
\begin{align}
\psi_{\lambda}(\boldsymbol{r},t)= & \sqrt{\frac{k_{\perp,c}}{2\pi\sigma_{z}\sigma_{\rho}}}i^{m+\lambda}J_{m+\lambda}(k_{\perp,c}\rho)e^{i(m+\lambda)\varphi}\exp\left[-\frac{\left(ct-z\cos\theta_{c}\right)^{2}}{4\sigma_{z}^{2}\cos^{2}\theta_{c}}+ik_{z,c}(z-ct/\cos\theta_{c})\right]\\
= & \sqrt{\frac{C_{0}}{2\pi V_p}}i^{m+\lambda}J_{m+\lambda}(k_{\perp,c}\rho)e^{i(m+\lambda)\varphi}\exp\left[-\frac{\left(ct-z\cos\theta_{c}\right)^{2}}{4\sigma_{z}^{2}\cos^{2}\theta_{c}}+ik_{z,c}(z-ct/\cos\theta_{c})\right],
\end{align}
where the constant $C_{0}=k_{z,c}\sigma_{z}=k_{\perp,c}\sigma_{\rho}\gg1$ describes the sharpness of the SAF, $V_{p}=\sigma_{z}\sigma_{\rho}^{2}$ characterizes the effective
volume of the photon pulse, and we have used the Jacobi-Anger expansion
\begin{align}
e^{iz\cos\theta} & =\sum_{n=-\infty}^{\infty}i^{n}J_{n}(z)e^{in\theta},\\
e^{-iz\cos\theta} & =e^{iz\cos(\theta+\pi)}=\sum_{n=-\infty}^{\infty}i^{n}J_{n}(z)e^{in(\theta+\pi)}=\sum_{n=-\infty}^{\infty}(-i)^{n}J_{n}(z)e^{in\theta},
\end{align}
with the $n$th Bessel function of the first kind 
\begin{equation}
J_{n}(z)=\frac{1}{2\pi}\int_{0}^{2\pi}e^{i(n\varphi-z\sin\varphi)}d\varphi.
\end{equation}

Then, we have
\begin{equation}
\left\langle n_{\xi\lambda}\right|\hat{\boldsymbol{s}}^{\rm obs}\left|n_{\xi\lambda}\right\rangle =s_{\lambda}\boldsymbol{e}_{\lambda}+s_{-\lambda}\boldsymbol{e}_{-\lambda}+\lambda s_{z}\boldsymbol{e}_{z}
\end{equation}
where $s_{-\lambda}=s_{\lambda}^{*}$,
\begin{align}
s_{\lambda}= & -in\hbar\times\frac{i\lambda}{\sqrt{2}}\sin\theta_{c}\left[\cos^{2}\frac{\theta_{c}}{2}\psi_{0}^{*}(\boldsymbol{r},t)\psi_{-\lambda}(\boldsymbol{r},t)+\sin^{2}\frac{\theta_{c}}{2}\psi_{\lambda}^{*}(\boldsymbol{r},t)\psi_{0}(\boldsymbol{r},t)\right]\\
= & -\frac{in\hbar C_{0}\sin\theta_{c}}{2\sqrt{2}\pi\sigma_{z}\sigma_{\rho}^{2}}e^{-i\lambda\varphi}\left[\cos^{2}\frac{\theta_{c}}{2}J_{m}(k_{\perp,c}\rho)J_{m-\lambda}(k_{\perp,c}\rho)+\sin^{2}\frac{\theta_{c}}{2}J_{m+\lambda}(k_{\perp,c}\rho)J_{m}(k_{\perp,c}\rho)\right]\exp\left[-\frac{\left(ct-z\cos\theta_{c}\right)^{2}}{2\sigma_{z}^{2}\cos^{2}\theta_{c}}\right],
\end{align}
and
\begin{align}
s_{z} & =-in\hbar\times i\left[\left|\psi_{-\lambda}(\boldsymbol{r},t)\right|^{2}\cos^{4}\frac{\theta_{c}}{2}-\left|\psi_{\lambda}(\boldsymbol{r},t)\right|^{2}\sin^{4}\frac{\theta_{c}}{2}\right]\\
 & =\frac{n\hbar C_{0}}{2\pi\sigma_{z}\sigma_{\rho}^{2}}\left\{ \left[J_{m-\lambda}(k_{\perp,c}\rho)\right]^{2}\cos^{4}\frac{\theta_{c}}{2}-\left[J_{m+\lambda}(k_{\perp,c}\rho)\right]^{2}\sin^{4}\frac{\theta_{c}}{2}\right\} \exp\left[-\frac{\left(ct-z\cos\theta_{c}\right)^{2}}{2\sigma_{z}^{2}\cos^{2}\theta_{c}}\right].
\end{align}
Using the relation $\boldsymbol{e}_{\pm\lambda}=(\boldsymbol{e}_{x}\pm i\lambda\boldsymbol{e}_{y})/\sqrt{2}$,
we can express the spin density vector as
\begin{align}
\left\langle n_{\xi\lambda}\right|\hat{\boldsymbol{s}}^{\rm obs}(\boldsymbol{r},t)\left|n_{\xi\lambda}\right\rangle  & =-s_{\varphi}\boldsymbol{e}_{x}\sin\lambda\varphi+\lambda s_{\varphi}\boldsymbol{e}_{y}\cos\lambda\varphi+\lambda s_{z}\boldsymbol{e}_{z} =\lambda\left(s_{\varphi}\boldsymbol{e}_{\varphi}+s_{z}\boldsymbol{e}_{z}\right),\label{eq:helical}
\end{align}
where
\begin{align}
s_{\varphi} & =\frac{n\hbar C_{0}\sin\theta_{c}}{2\pi\sigma_{z}\sigma_{\rho}^{2}}\left[\cos^{2}\frac{\theta_{c}}{2}J_{m}(k_{\perp,c}\rho)J_{m-\lambda}(k_{\perp,c}\rho)+\sin^{2}\frac{\theta_{c}}{2}J_{m+\lambda}(k_{\perp,c}\rho)J_{m}(k_{\perp,c}\rho)\right]\exp\left[-\frac{\left(ct-z\cos\theta_{c}\right)^{2}}{2\sigma_{z}^{2}\cos^{2}\theta_{c}}\right]
\end{align}
The spin density of a coherent-state Bessel pulse can be evaluated similarly. 

\subsection*{Energy density}
To visualise the pulse profile, we give the energy density of the pulse, which is characterized by
\begin{align}
\hat{I}_{E}(\boldsymbol{r},t) = & \frac{1}{2}\left[\varepsilon_{0}\hat{\boldsymbol{E}}_{\perp}^{2}(\boldsymbol{r},t)+\frac{1}{\mu_{0}}\hat{\boldsymbol{B}}^{2}(\boldsymbol{r},t)\right]\\
  \approx & \frac{\hbar}{4(2\pi)^{3}}{\sum_{\lambda'\lambda''}}\int d^{3}k'\int d^{3}k''\sqrt{\omega_{\boldsymbol{k}'}\omega_{\boldsymbol{k}''}} \nonumber \\
 & \left\{(1+\lambda'\lambda'')\hat{a}_{\boldsymbol{k}',\lambda'}^{\dagger}\hat{a}_{\boldsymbol{k}'',\lambda''}e^{-i[(\boldsymbol{k}'-\boldsymbol{k}'')\cdot\boldsymbol{r}-(\omega_{\boldsymbol{k}'}-\omega_{\boldsymbol{k}''})t]}[\boldsymbol{e}^{*}(\boldsymbol{k}',\lambda')\cdot\boldsymbol{e}(\boldsymbol{k}'',\lambda'')]+{\rm h.c.}\right\}, 
\end{align}
where we have neglected the zero-point energy.

In the quasi-single-frequency limit with $\omega_{\boldsymbol{k}'}=\omega_{\boldsymbol{k}''}$ and $\theta_{\boldsymbol{k}'}=\theta_{\boldsymbol{k}''}=\theta_c$, its mean value is approximated
as
\begin{align}
\left\langle n_{\xi m\lambda}\right|\hat{I}_{E}(\boldsymbol{r},t)\left|n_{\xi m\lambda}\right\rangle
& \approx  n\hbar\omega_{c}\frac{1}{(2\pi)^3}\int d^{3}k'\int d^{3}k''\left[\xi_{m\lambda}^{*}(\boldsymbol{k}')\xi_{m\lambda}(\boldsymbol{k}'')\right]e^{-i[(\boldsymbol{k}'-\boldsymbol{k}'')\cdot\boldsymbol{r}]}[\boldsymbol{e}^{*}(\boldsymbol{k}',\lambda)\cdot\boldsymbol{e}(\boldsymbol{k}'',\lambda)]\\
 & \approx n\hbar\omega_{c}\left[\left|\psi_{-\lambda}\right|^{2}\cos^{4}\frac{\theta_{c}}{2}+\left|\psi_{\lambda}\right|^{2}\sin^{4}\frac{\theta_{c}}{2}+\frac{1}{2}\left|\psi_{0}\right|^{2}\sin^{2}\theta_{c}\right].
\end{align}

\vspace{0.5cm}
\noindent\textbf{Supplementary Note 6: Quantum correlation of spin density}
\vspace{0.5cm}

In this section, we show how to evaluate the quantum correlations of the photon spin density. Due to the vector nature of the spin density, the full two-point correlation should be characterized by a $3\times 3$ correlation matrix
\begin{equation}
G(\boldsymbol{r},t;\boldsymbol{r}',t')=\left[\begin{array}{ccc}
\left\langle \hat{s}_{x}^{{\rm obs}}(\boldsymbol{r},t)\hat{s}_{x}^{{\rm obs}}(\boldsymbol{r}',t')\right\rangle  & \left\langle \hat{s}_{x}^{{\rm obs}}(\boldsymbol{r},t)\hat{s}_{y}^{{\rm obs}}(\boldsymbol{r}',t')\right\rangle  & \left\langle \hat{s}_{x}^{{\rm obs}}(\boldsymbol{r},t)\hat{s}_{z}^{{\rm obs}}(\boldsymbol{r}',t')\right\rangle \\
\left\langle \hat{s}_{y}^{{\rm obs}}(\boldsymbol{r},t)\hat{s}_{x}^{{\rm obs}}(\boldsymbol{r}',t')\right\rangle  & \left\langle \hat{s}_{y}^{{\rm obs}}(\boldsymbol{r},t)\hat{s}_{y}^{{\rm obs}}(\boldsymbol{r}',t')\right\rangle  & \left\langle \hat{s}_{y}^{{\rm obs}}(\boldsymbol{r},t)\hat{s}_{z}^{{\rm obs}}(\boldsymbol{r}',t')\right\rangle \\
\left\langle \hat{s}_{z}^{{\rm obs}}(\boldsymbol{r},t)\hat{s}_{x}^{{\rm obs}}(\boldsymbol{r}',t')\right\rangle  & \left\langle \hat{s}_{z}^{{\rm obs}}(\boldsymbol{r},t)\hat{s}_{y}^{{\rm obs}}(\boldsymbol{r}',t')\right\rangle  & \left\langle \hat{s}_{z}^{{\rm obs}}(\boldsymbol{r},t)\hat{s}_{z}^{{\rm obs}}(\boldsymbol{r}',t')\right\rangle 
\end{array}\right].
\end{equation}
Here, we only concern the equal-time correlator $\left\langle \hat{s}^{\rm obs}_{z}(\boldsymbol{r},t)\hat{s}^{\rm obs}_{z}(\boldsymbol{r}',t)\right\rangle$.
To simplify the calculation, we take the paraxial approximation,
i.e., $\theta_{c}\approx0$. For the contact term below, we use the ideal local-continuum approximation; a finite paraxial bandwidth replaces $\delta^{3}(\boldsymbol r-\boldsymbol r')$ by the corresponding band-limited kernel. Then, the two-point spin-density correlations for a Fock-state and coherent-state pulse are given by
\begin{align}
 & \left\langle n_{\xi\lambda}\right|\hat{s}^{\rm obs}_{z}(\boldsymbol{r},t)\hat{s}^{\rm obs}_{z}(\boldsymbol{r}',t)\left|n_{\xi\lambda}\right\rangle \nonumber \\
\approx 
& \frac{\hbar^{2}}{(2\pi)^{6}}\int d^{3}k_{1}\int d^{3}k_{2}\int d^{3}k_{3}\int d^{3}k_{4}e^{-i[(\boldsymbol{k}_{1}-\boldsymbol{k}_{2})\cdot\boldsymbol{r}+(\boldsymbol{k}_{3}-\boldsymbol{k}_{4})\cdot\boldsymbol{r}'-(\omega_{1}-\omega_{2}+\omega_{3}-\omega_{4})t]}e^{i\lambda(\varphi_{k_{1}}-\varphi_{k_{2}}+\varphi_{k_{3}}-\varphi_{k_{4}})} \\
 & \times\left\langle n_{\xi\lambda}\right|\hat{a}_{\boldsymbol{k}_{1},\lambda}^{\dagger}\hat{a}_{\boldsymbol{k}_{3},\lambda}^{\dagger}\hat{a}_{\boldsymbol{k}_{2},\lambda}\hat{a}_{\boldsymbol{k}_{4},\lambda}+\delta(\boldsymbol{k}_{2}-\boldsymbol{k}_{3})\hat{a}_{\boldsymbol{k}_{1},\lambda}^{\dagger}\hat{a}_{\boldsymbol{k}_{4},\lambda}\left|n_{\xi\lambda}\right\rangle \\
= & \hbar^{2}\left[n(n-1)\left|\psi_{-\lambda}(\boldsymbol{r},t)\right|^{2}\left|\psi_{-\lambda}(\boldsymbol{r}',t)\right|^{2}+\delta(\boldsymbol{r}-\boldsymbol{r}')n\left|\psi_{-\lambda}(\boldsymbol{r},t)\right|^{2}\right],
\end{align}
and 
\begin{align}
& \left\langle \alpha_{\xi\lambda}\right|\hat{s}^{\rm obs}_{z}(\boldsymbol{r},t)\hat{s}^{\rm obs}_{z}(\boldsymbol{r}',t)\left|\alpha_{\xi\lambda}\right\rangle \nonumber \\
\approx & \frac{\hbar^{2}}{(2\pi)^{6}}\int d^{3}k_{1}\int d^{3}k_{2}\int d^{3}k_{3}\int d^{3}k_{4}e^{-i[(\boldsymbol{k}_{1}-\boldsymbol{k}_{2})\cdot\boldsymbol{r}+(\boldsymbol{k}_{3}-\boldsymbol{k}_{4})\cdot\boldsymbol{r}'-(\omega_{1}-\omega_{2}+\omega_{3}-\omega_{4})t]}e^{i\lambda(\varphi_{k_{1}}-\varphi_{k_{2}}+\varphi_{k_{3}}-\varphi_{k_{4}})} \\
 & \times\left\langle \alpha_{\xi\lambda}\right|\hat{a}_{\boldsymbol{k}_{1},\lambda}^{\dagger}\hat{a}_{\boldsymbol{k}_{3},\lambda}^{\dagger}\hat{a}_{\boldsymbol{k}_{2},\lambda}\hat{a}_{\boldsymbol{k}_{4},\lambda}+\delta(\boldsymbol{k}_{2}-\boldsymbol{k}_{3})\hat{a}_{\boldsymbol{k}_{1},\lambda}^{\dagger}\hat{a}_{\boldsymbol{k}_{4},\lambda}\left|\alpha_{\xi\lambda}\right\rangle \\
= & \hbar^{2}\left[\bar{n}^{2}\left|\psi_{-\lambda}(\boldsymbol{r},t)\right|^{2}\left|\psi_{-\lambda}(\boldsymbol{r}',t)\right|^{2}+\delta(\boldsymbol{r}-\boldsymbol{r}')\bar{n}\left|\psi_{-\lambda}(\boldsymbol{r},t)\right|^{2}\right]
\end{align}
Here, we see that the Poisson and sub-Poisson statistics automatically enter the quantum spin-density correlations.

We note that the delta function $\delta(\boldsymbol{r}-\boldsymbol{r}')$ in the correlator will not lead to any diverging effect, because a practical probe always measures
the averaged photon spin density over a finite volume instead of the true single-point spin density. We now consider a finite detector volume $V_d$ over which the pulse profile is approximately constant; the following $1/V_d$ terms use the ideal local-delta approximation.
For a Fock-state Bessel pulse, the averaged photon spin density detected
by this probe is given by
\begin{equation}
\left\langle n_{\xi\lambda}\right|\frac{1}{V_{d}}\int_{V_{d}(\boldsymbol r)}d^{3}u\,\hat{s}^{\rm obs}_{z}(\boldsymbol{u},t)\left|n_{\xi\lambda}\right\rangle \approx \lambda n\hbar\left|\psi_{-\lambda}(\boldsymbol{r},t)\right|^{2}.
\end{equation}
The corresponding spin correlation is given by
\begin{gather*}
\left\langle n_{\xi\lambda}\right|\frac{1}{V_{d}^{2}}\int_{V_d(\boldsymbol r)}d^{3}u\,\hat{s}^{\rm obs}_{z}(\boldsymbol{u},t)\int_{V_d(\boldsymbol r')}d^{3}v\,\hat{s}^{\rm obs}_{z}(\boldsymbol{v},t)\left|n_{\xi\lambda}\right\rangle \\[-1pt]
\approx \begin{cases}
\hbar^{2}n(n-1)\left|\psi_{-\lambda}(\boldsymbol{r},t)\right|^{2}\left|\psi_{-\lambda}(\boldsymbol{r}',t)\right|^{2}, & {\rm for}\ \boldsymbol{r}\neq\boldsymbol{r}'\\
\hbar^{2}n(n-1)\left|\psi_{-\lambda}(\boldsymbol{r},t)\right|^{4}+\frac{n\hbar^{2}}{V_{d}}\left|\psi_{-\lambda}(\boldsymbol{r},t)\right|^{2}, & {\rm for}\ \boldsymbol{r}=\boldsymbol{r}'
\end{cases}
\end{gather*}
Then, the uncertainty of the spin density in $z$-direction is given by
\begin{equation}
\Delta\hat{s}^{\rm obs}_{z} \approx\hbar\sqrt{n}\sqrt{\frac{1}{V_{d}}\left|\psi_{-\lambda}(\boldsymbol{r},t)\right|^{2}-\left|\psi_{-\lambda}(\boldsymbol{r},t)\right|^{4}}.
\end{equation}
Usually, the second term will be very small, i.e., the ratio of the square roots of the first and second terms is of order $\sqrt{V_{p}/V_{d}}\gg\sqrt{\lambda^3_c/V_{d}}\gg 1$.

Similarly for a coherent-state pulse, we have
\begin{equation}
\left\langle \alpha_{\xi\lambda}\right|\frac{1}{V_{d}}\int_{V_{d}(\boldsymbol r)}d^{3}u\,\hat{s}^{\rm obs}_{z}(\boldsymbol{u},t)\left|\alpha_{\xi\lambda}\right\rangle \approx \lambda\bar{n}\hbar\left|\psi_{-\lambda}(\boldsymbol{r},t)\right|^{2}.
\end{equation}
The corresponding spin correlation is given by
\begin{gather*}
\left\langle \alpha_{\xi\lambda}\right|\frac{1}{V_{d}^{2}}\int_{V_d(\boldsymbol r)}d^{3}u\,\hat{s}^{\rm obs}_{z}(\boldsymbol{u},t)\int_{V_d(\boldsymbol r')}d^{3}v\,\hat{s}^{\rm obs}_{z}(\boldsymbol{v},t)\left|\alpha_{\xi\lambda}\right\rangle \\[-1pt]
\approx \begin{cases}
\hbar^{2}\bar{n}^2\left|\psi_{-\lambda}(\boldsymbol{r},t)\right|^{2}\left|\psi_{-\lambda}(\boldsymbol{r}',t)\right|^{2}, & {\rm for}\ \boldsymbol{r}\neq\boldsymbol{r}'\\
\hbar^{2}\bar{n}^2\left|\psi_{-\lambda}(\boldsymbol{r},t)\right|^{4}+\frac{\bar{n}\hbar^{2}}{V_{d}}\left|\psi_{-\lambda}(\boldsymbol{r},t)\right|^{2}, & {\rm for}\ \boldsymbol{r}=\boldsymbol{r}'
\end{cases}
\end{gather*}
Then, the corresponding uncertainty of the spin density in $z$-direction is given by
\begin{equation}
\Delta\hat{s}^{\rm obs}_{z} \approx\hbar\sqrt{\frac{\bar{n}}{V_{d}}}\left|\psi_{-\lambda}(\boldsymbol{r},t)\right|.
\end{equation}

\end{document}